\documentclass[11pt]{article}
\date{\today}
\pdfoutput=1
\usepackage{jcappub}
\usepackage[T1]{fontenc}
\usepackage{subfig}
\usepackage{caption}
\usepackage{lineno,hyperref}
\usepackage{amsmath}
\usepackage{amssymb}
\usepackage{mathtools}
\usepackage{tablefootnote}
\usepackage{soul}
\usepackage{ulem}
\usepackage{color}
\usepackage{float}
\usepackage{makecell}
\usepackage{lmodern}
\DeclareFontFamily{OMX}{lmex}{}
\DeclareFontShape{OMX}{lmex}{m}{n}{<-> lmex10}{}
\usepackage[usenames,dvipsnames,svgnames,table]{xcolor}
\usepackage{scalerel}[2016-12-29]
\def\scaleint#1{\vcenter{\hbox{\scaleto[3ex]{\displaystyle\int}{#1}}}}
\def\bs{\mkern-12mu}

\newcommand{\CE}{classical extension}
\newcommand{\EMs}{elaborate models}

\usepackage{siunitx}
\title{Review of Hubble tension solutions with new SH0ES and SPT-3G data}
\author[a]{Ali Rida Khalife,}
\emailAdd{ridakhal@iap.fr}
\author[b]{Maryam Bahrami Zanjani,}
\emailAdd{bahrami.zanjani.m@gmail.com}
\author[a]{Silvia Galli,}
\emailAdd{gallis@iap.fr}
\author[b]{Sven Günther,}
\emailAdd{sven.guenther@rwth-aachen.de}
\author[b]{Julien Lesgourgues,}
\emailAdd{lesgourg@physik.rwth-aachen.de}
\author[a]{Karim Benabed.}
\emailAdd{benabed@iap.fr}

\affiliation[a]{Sorbonne Universit\'e, CNRS, UMR 7095, Institut d’Astrophysique de Paris,
98 bis bd Arago, 75014 Paris, France}
\affiliation[b]{Institute for Theoretical Particle Physics and Cosmology  (TTK),
RWTH Aachen University, Sommerfeldstr. 16, D-52056 Aachen, Germany.}

\abstract{
     We present an updated analysis of eleven cosmological models that may help reduce the Hubble tension, which now reaches the $6\sigma$ level when considering the latest SH0ES measurement versus recent CMB and BAO data, assuming $\Lambda$CDM.
      Specifically, we look at five classical extensions of $\Lambda$CDM (with massive neutrinos, spatial curvature, free-streaming or self-interacting relativistic relics, or dynamical dark energy) and six elaborate models featuring either a time-varying electron mass, early dark energy or some non-trivial interactions in the neutrino sector triggered by a light Majoron. We improve over previous works in several ways. We include the latest data from the South Pole Telescope as well as the most recent measurement of the Hubble rate by the SH0ES collaboration. We treat the summed neutrino mass as a free parameter in most of our models, which reveals interesting degeneracies and constraints. We define additional metrics to assess the potential of a model to reduce or even solve the Hubble tension. We validate an emulator that uses active learning to train itself during each parameter inference run for any arbitrary model. We find that the time-varying electron mass and the Majoron models are now ruled out at more than $3\sigma$. Models with a time-varying electron mass and spatial curvature or with early dark energy reduce the tension to $1.0-2.9\sigma$. Nevertheless, none of the models considered in this work is favored with enough statistical significance to become the next concordance model of Cosmology.
     }

\begin{document}

\begin{flushleft}
TTK-23-36
\end{flushleft}

\maketitle
\flushbottom

\section{Introduction}
\label{Sec:Intro}

Since its appearance almost a decade ago, the Hubble tension~\cite{Planck:2013,Verde1,VerdeTreuRiess} has become one of the most debated topics in modern Cosmology~\cite{In_Realm,H0_Olympics} (see figures 1 of~\cite{Cosmology_Crossroad,HT_History} for the evolution of Hubble tension over the years). Simply stated, this tension arises from a discrepancy between values of the current expansion rate of the Universe, the Hubble parameter ($H_0$), inferred either indirectly and in a model-dependent way from the angular scale of a few standard rulers, or directly from the luminosity of standard candles. 

The former method mainly includes observations of the sound horizon angular scales in Cosmic Microwave Background (CMB) and Baryon Acoustic Oscillation (BAO) data.\footnote{One can also use another standard ruler, the horizon at matter-radiation equality~\cite{Philcox:2022sgj}. This provides results consistent with those from the sound horizon.} Assuming the minimal cosmological model, $\Lambda$CDM, the Planck collaboration reports $H_0 = 67.36 \pm 0.54\,$km/s/Mpc~at the 68\% confidence level (CL)~\cite{Planck2018}.\footnote{This comes from the analysis of Planck Release 3 (PR3) data with temperature, polarization, and lensing information (TT,TE,EE+lowE+lensing). Although our work is based on PR3, we note that a recent analysis of PR4 by ref.~\cite{Tristram:2023haj} yields a similar but slightly more constraining result, $H_0 = 67.66 \pm 0.49\,$km/s/Mpc (see also ref.~\cite{George_Paper}).} However, it is crucial to note that similar values have been inferred by other CMB~\cite{ACT_DR6,SPT-3G:2021,SPT-3G:2022} and non-CMB~\cite{Pantheon,SDSS_DR12,SDSS_DR16,BBN1,BBN2,BBN_BAO_Nils,Philcox:2022sgj} probes of the early Universe, showing that the tension subsists even when excluding Planck.

On the other hand, the latter method relies mainly on the luminosity of type-Ia supernovae (SNIa) calibrated by Cepheids using the distance ladder technique~\cite{Distance_Ladder2}. For this measurement, the main contender is the SH0ES collaboration~\cite{Riess:2019cxk,Riess:2020fzl,Riess:2021jrx}, which recently presented an improved measurement of $H_0 = 73.29 \pm 0.90$ km/s/Mpc~\cite{New_H0}, in 5.6$\sigma$ tension with the Planck measurement. Like for the measurement of the angular scale of standard rulers, other late universe observations are hinting at similar conclusions, although with much lower significance~\cite{H0LICOW,MIRA,H0_SNII,TRGB1,TRGB2,Distance_Ladder1,TRGB3,LIGO_H0,CosmoGraphic_H0}.


In order to resolve this tension, one might consider two different approaches. The first reassesses the data and its analysis in an attempt to uncover some hidden systematics. A great deal of work has been done along this direction, both for the angular scale of standard rulers (see section 6 of ref.~\cite{Planck_systematics} and references therein) and for distance ladders~\cite{Mortsell:2021tcx,systematics1,systematics2,Efstathiou_systematics,systematics4,systematics5,systematics6,Systematics3}. However, these efforts did not really reduce or resolve the tension significantly, and some re-analyses even increased it further~\cite{New_H0}. Hence, even though the possibility of hidden systematics is not completely excluded, one can attempt the second approach: looking for alternative models to $\Lambda$CDM. 

The minimal $\Lambda$CDM model is extremely successful at describing many observations with just 6 parameters. Nevertheless, this model has a few features that make it incomplete, at least from a theoretical cosmology perspective. These include the fine-tuning and origin of the cosmological constant $\Lambda$, the coincidence problem, the nature of Dark Matter (DM), and the origin of cosmological inflation (see~\cite{EUcapt_White_Paper,Beyond_LCDM,MG_Cosmology} for reviews). Furthermore, since measurements of the angular scale of the sound horizon infer the value of $H_0$ under the assumption of a given cosmological model, i.e. indirectly, one should naturally consider the potential of different models to solve the Hubble tension. 

To achieve this goal, one must search for a model in which the extremely precise determination of the angular scale of the sound horizon by CMB and BAO data remains compatible with higher values of $H_0$ than in the $\Lambda$CDM model -- or even, ideally, predicts a higher value of $H_0$ than this model. In natural units, where $\hbar=c=1$, the sound horizon angular scale is given by
\begin{equation}
        \theta_s = \frac{r_s(z_\mathrm{dec})}{r_A(z_\mathrm{obs})}=\frac{\scaleint{9ex}_{\bs z_\mathrm{dec}}^{\infty}\bigg[3\bigg (1+\frac{3\rho_b}{4\rho_{\gamma}}\bigg)\bigg]^{-1/2}\bigg[\frac{8\pi G}{3}\Sigma_i\rho_i\bigg]^{-1/2}
        dz}{H_0^{-1}\sin_K\bigg[\scaleint{6ex}_{\bs 0}^{z_\mathrm{obs}}\bigg (\Sigma_i\Omega_i (z)\bigg)^{-1/2} dz\bigg]},
        \label{Eq:Theta}
\end{equation}  
where $r_s(z_\mathrm{dec})$ is the comoving sound horizon at the time of decoupling (for CMB observations, at the time of photon decoupling, and for BAO observations, at the baryon drag time);
$r_A(z_\mathrm{obs})$ is the comoving angular diameter distance to observations (for CMB observations, to the time of photon decoupling, and for BAO observations, to the mean redshift of the galaxies used in the measurement);
$G$ is Newton's constant;
$\rho_i$ is the energy density of species $i$, with $\rho_b$ and $\rho_{\gamma}$ being that of baryons and photons, respectively; finally, $\Omega_i=8\pi G\rho_i/ (3H_0^2)$ for species $i$, and spatial curvature is parameterized by $\Omega_K=-K/H_0^2$ (see eq.(2.39) of~\cite{Dodelson} for the definition of $\sin_K$ above).

The angle $\theta_s$ is extremely well constrained, especially by CMB  observations ($100\,\theta_s= 1.04075 \pm 0.00028$ at the 68\%CL~\cite{SPT-3G:2022}). Therefore, we can see from the last part of eq.~\eqref{Eq:Theta} that, if $H_0$ is to match the late Universe value (without shifting $100\,\theta_s$ by more than a few $\sigma$'s), one must change some, or all, of the other terms. This can be done in two ways: by modifying the ingredients of the cosmological model before recombination ($z \geq z_\mathrm{dec}$) or after recombination  ($z \leq z_\mathrm{obs}$). An example of the former solution consists in increasing the energy budget of relativistic species in that epoch, i.e. increase $\sum_i \rho_i$ in the numerator of eq.~\eqref{Eq:Theta}, enough to have $H_0\sim73$ km/s/Mpc. In this case, one should take care of avoiding a conflict with Big Bang Nucleosynthesis (BBN) bounds on the relativistic energy density at even earlier times -- some of the models considered here address this problem successfully, but in this work, for simplicity, we will not enter into further considerations concerning BBN. As an example of a post-recombination solution, one may assume a decreasing amount of Dark Energy  (DE) in the late universe by making its equation of state (EoS) parameter, $w_{\text{DE}}$, dynamical~\cite{CPL1,CPL2} (we define
$w_{\text{DE}}=p_{\text{DE}}/\rho_{\text{DE}}$, where $p_{\text{DE}}$ and $\rho_{\text{DE}}$ are the pressure and energy density of DE, respectively).\footnote{It should be noted, however, that this particular model was shown not to solve the Hubble tension~\cite{CPL_No_1,CPL_NO_2,SDSS_DR12}. We mention it here just as an illustration.} In principle one can always modify both eras, as we shall see later for two of the models considered here. This option is further discussed on a general basis in ref.~\cite{Vagnozzi:2023nrq}.

The aforementioned examples are a mere sample of a plethora of models that have been put forward to resolve the Hubble tension, with either pre- or post-recombination modifications of $\Lambda$CDM. Many other models have been extensively assessed (see refs.~\cite{H0_Olympics,In_Realm,Escudero:2022rbq,No-goGuide1} for non-exhaustive reviews) and sometimes ruled out as viable solutions.  However, given that a final consensus has not been reached yet, it is important to build upon these previous analyses with new, state-of-the-art data, to better constrain the remaining models. This is precisely the goal of the current work.

We present updated constraints on twelve non-minimal cosmologies, including $\Lambda$CDM for comparison. We divide these models into \CE{}s of $\Lambda$CDM, i.e. with one or two parameters that have solid physical motivations and/or are interesting to confront to the data independently of the Hubble tension; and more elaborate models that have been previously suggested as promising solutions to the tension in ref.~\cite{H0_Olympics}.
The latter models include the varying electron mass scenario  ($m_e$)~\cite{VarMe1,VarMe2015,VarMe2018,Planck_VarMe}  and some of its extensions~\cite{VarMe_Omk}, Early Dark Energy (EDE)~\cite{EDE1,EDE2,EDE3}, and a Majoron-inspired model~\cite{Majoron1,Majoron2,Majoron3}. In most cases, we generalise over previous analyses by treating the summed neutrino mass $\Sigma m_{\nu}$ as a free parameter. This is a more physically motivated choice compared to fixing the mass to the minimum value allowed by laboratory experiments.\footnote{ We do not vary $\Sigma m_{\nu}$ for EDE and the Majoron models. The former already has many free parameters and adding $\Sigma m_{\nu}$ would make the convergence of the fitting algorithm prohibitively slow. In the Majoron case, the only theory code available at the moment neglects the small effect of neutrino masses.} It also allows us to investigate possible degeneracies between the neutrino mass and other cosmological parameters in extended models. After presenting a concise summary of our main results in section~\ref{Sec: Summary}, we briefly elaborate on the twelve models included in this work (and their physical parameters) in section~\ref{Sec:Models}. 
 
We confront each of these models with distinct data sets and their combinations,  including recent Baryon Acoustic Oscillations (BAO)~\cite{SDSS_6dF,SDSS_DR7,SDSS_DR12,SDSS_DR16}, SNIa~\cite{Pantheon} and CMB~\cite{Planck2018,ACT1,ACT2,SPT-3G:2021,SPT-3G:2022} data, as presented in section~\ref{Sec:Data and Method}. In our analysis, we underline the role and the constraining power of SPT-3G 2018 data~\cite{SPT-3G:2021,SPT-3G:2022}  compared to other CMB ones. 
For each data set, we assess the ability of the various models to provide an explanation for the Hubble tension through a variety of metrics discussed in section~\ref{Sec:Data and Method}. In the same section, we show that we have been able to speed up our pipeline for Bayesian parameter extraction and for likelihood maximization by a large factor without biasing the results, thanks to a new emulator~\cite{G_nther_2022} that uses an active learning scheme and requires no prior training.

Section~\ref{Sec:Results} contains an extensive discussion of our results. We conclude our work in section~\ref{Sec:Conclusion}. In appendix~\ref{APP: Pipeline_Plus} we give details on our parameter inference pipeline, emulator, and minimization technique.


\section{Results summary}
\label{Sec: Summary}

We fitted our models to many different combinations of data from CMB (Planck, SPT-3G 2018, ACT-DR4), BAO (6dFGS, BOSS, eBOSS), and SNIa (Pantheon) observations. However, to summarize our main results and compare the ability of different models to explain (or at least reduce) the Hubble tension, we focus on the combination of Planck, SPT-3G 2018, BAO, and SNIa data.

Indeed, the earlier work of ref.~\cite{H0_Olympics} considered the combination of Planck (TT, TE, EE+lowE+lensing), BAO (6dFGS, DR7, DR12), and SNIa (Pantheon) data as their baseline. Ref.~\cite{H0_Olympics} also performed additional tests with data from ACT-DR4 and concluded that it did not impact significantly the main conclusions, except for EDE. Indeed, it was shown in previous works~\cite{ACT_EDE,Recent_EDE_SPT,Ups_Downs_EDE} that there is a preference for EDE from ACT-DR4 data. We will return to this point in section~\ref{Sec:Results}. Since one of our main objectives is to highlight the impact of SPT-3G 2018 data, we do add it to the previously used baseline data set. We also check the role of ACT-DR4 data (on top of SPT-3G 2018) in section~\ref{Sec:Results}, but we do not include it in our baseline, since its impact has already been studied.
Note that our baseline also differs from that of ref.~\cite{H0_Olympics} through the inclusion of some recent measurements of the BAO scale by eBOSS DR16 (see section~\ref{subsec:data} for more details and references on the observational data used in this work).

\begin{figure}[hbt!]
        \hspace{-0.5cm} 
        \includegraphics[width=16cm,angle=0]{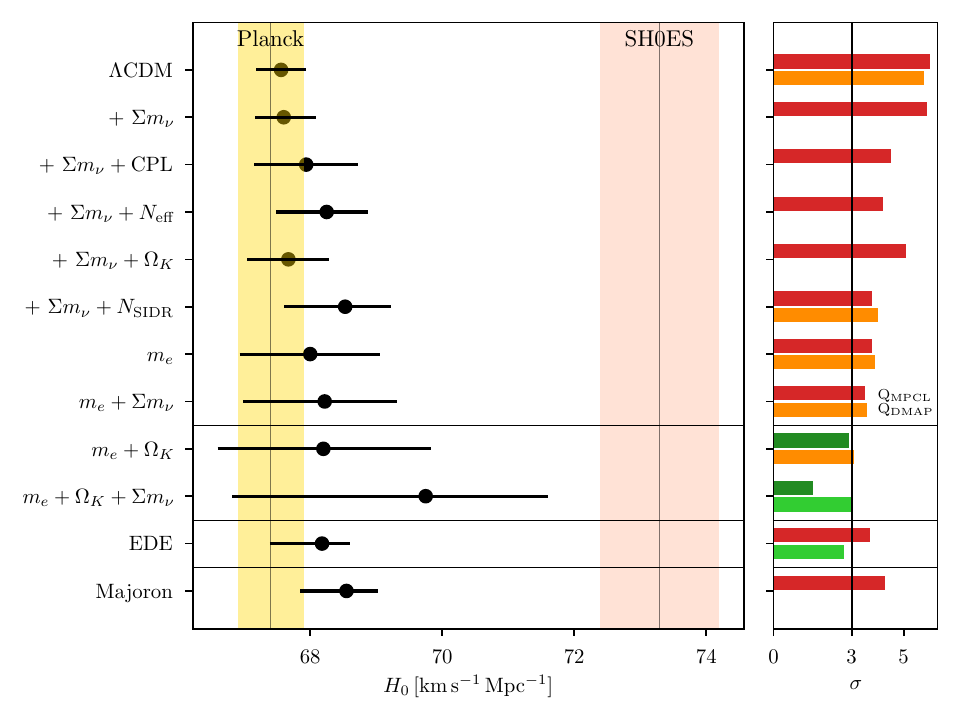}
        \caption{A graphical representation of the first four columns of Table~\ref{Table:Summary}. {\it Left}: The black dots and their intervals correspond to the mean value of $H_0$ and its 68\%CL, respectively, for each model. The gold band is the 68\% CL on $H_0$ from Planck~\cite{Planck2018}, while the light-salmon one corresponds to that of SH0ES~\cite{New_H0}. {\it Right}: The top horizontal bar of each model correspond to $Q_\mathrm{MPCL}$, colored in red  when surpassing the $3\sigma$ level, and dark green otherwise. The bottom horizontal bar, when applicable, corresponds to $Q_\mathrm{DMAP}$, colored in orange when surpassing the $3\sigma$ level, and light green otherwise }
        \label{fig:Graphical_Summary}
\end{figure}

Our main results for the different models and the baseline data set are presented in Table~\ref{Table:Summary} and summarized graphically in Figure~\ref{fig:Graphical_Summary}. The second column provides the mean value, best-fit value, and 68\%CL credible intervals for $H_0$ when each model is fitted to the baseline data set. The next columns of Table~\ref{Table:Summary} refer to several complementary approaches to quantify the level of the Hubble tension. Two of them have already been used in~\cite{H0_Olympics}: the ``Difference of the Maximum A Posteriori'' (DMAP) yields a number of $\sigma$'s denoted as $Q_\mathrm{DMAP}$ and listed in the the 4$^\mathrm{th}$ column, and the ``Akaike Information Criterion (AIC) with SH0ES'' yields a number denoted as $\Delta$AIC$_\mathrm{w/}$ and provided in the 8$^\text{th}$ column. However, we improve over the previous tests in two ways. First, we replace the Gaussian Tension (GT) metric of ref.~\cite{H0_Olympics} by its generalisation to the case of non-Gaussian posteriors, that we call the ``Marginalised Posterior Compatibility Level'' (MPCL) and translate into a number $Q_\mathrm{MPCL}$ of $\sigma$'s listed in the 3$^\mathrm{d}$ column (see section~\ref{sec:metrics}). Second, we also compute the ``AIC without SH0ES'', $\Delta$AIC$_\mathrm{w/o}$, reported in the 6$^\text{th}$ column. 

The meaning of these various metrics can be summarized as follows. Both $Q_\mathrm{MPCL}$ and $Q_\mathrm{DMAP}$ assess the compatibility between the value of $H_0$ predicted either by the baseline data set or by the SH0ES measurement, under the assumption of a given cosmological model. The former evaluates this level of compatibility in a Bayesian way (depending on priors and volume effects) and the latter in a frequentist way (depending solely on maximum likelihoods, or equivalently on best-fit $\chi^2$). Next, the $\Delta$AIC$_\mathrm{w/o}$ metric states whether an extended model is preferred over $\Lambda$CDM given the baseline data set only, and $\Delta$AIC$_\mathrm{w/}$ addresses the same question given the baseline plus SH0ES data set. We further elaborate on these parameters and their definitions in section~\ref{sec:metrics}. Note that the MPCL metric can be quickly inferred from the Monte Carlo Markov Chains (MCMC), while the DMAP and AIC metrics require minimizations, which can be numerically expensive (despite the speed-up resulting from the new method described in section \ref{subsec:minimization}, which uses an emulator for this purpose). For this reason, we compute the DMAP and AIC metrics only for $\Lambda$CDM and for extended models that have $Q_\mathrm{MPCL}\leq 4\sigma$.

    \begin{table}[ht!]
        \resizebox{\textwidth}{!}{%
            \centering
            \renewcommand{\arraystretch}{1.5}
            \begin{tabular} {l |c c c | c c | c c}
             \multicolumn{2}{c}{ } & & & \multicolumn{2}{c|}{w/o SH0ES} &  \multicolumn{2}{c}{w/ SH0ES}\\
            \hline
            Models &  $H_0$(km/s/Mpc) & Q$_\mathrm{MPCL} (\sigma)$ & Q$_\mathrm{DMAP} (\sigma)$ & $\Delta \chi^2$ & $\Delta$AIC & $\Delta \chi^2$ & $\Delta$AIC\\
            \hline
            {$\Lambda$CDM} & $67.56(67.58)^{+0.38}_{-0.38}$ & 6.0 & 5.8 & 0 & 0 & 0 & 0 \\
            
            {+$\Sigma m_{\nu}$} & $67.60(67.01)^{+0.49}_{-0.43}$ & 5.9 & --- & --- & --- & --- & --- \\
            
            {+$\Sigma m_{\nu}$+CPL} & $67.94(67.89) ^{+0.78}_{-0.79}$ & 4.5 & --- & --- & --- & --- & ---\\
            
            {+$\Sigma m_{\nu} +$ N${}_{\text{eff}}$} & $68.25(67.45)^{+0.62}_{-0.76}$ & 4.2 & --- & --- & --- & --- & ---\\
            
            {+$\Sigma m_{\nu}+ \Omega_{K}$} & $67.67(66.88)^{+0.62}_{-0.62}$ & 5.1 & ---  & --- & --- & --- & ---\\

            {+$\Sigma m_{\nu}$+ N$_{\text{SIDR}}$} & $68.53(69.06)^{+0.69}_{-0.92}$ & 3.8 & 4.0 & -0.1 & 3.9 & -17.1 &  -13.1 \\
            
            {$m_e$} & $68.00(68.03)^{+1.06}_{-1.07}$ & 3.8 & 3.9 & 0.0 & 2.0 & -18.0 & -16.0 \\
            
            {$m_e$+$\Sigma m_{\nu}$} & 68.22(67.70)$^{+1.09}_{-1.23}$ & 3.5 & 3.6 & -1.0 & 3.1 & -21.6 & -17.6 \\
            
            {$m_e$ + $\Omega_K$} & 68.20(67.42)$^{+1.63}_{-1.60}$ & \bf 2.9 & 3.1 & -1.0 & 3.0 & -24.7 & -20.7\\
            
            {$m_e$ + $\Omega_K$ + $\Sigma m_{\nu}$}& $69.75(67.75)^{+1.85}_{-2.93}$ & \bf 1.5 & \bf 3.0 & -0.9 & 5.1 & -25.6 & -19.8 \\
            
            {EDE} & 68.18(68.55)$^{+0.42}_{-0.79}$ & 3.7 & \bf 2.7 & -4.6 & 1.4 & -31.1 & -25.1 \\
            
            {Majoron} & $68.55(68.08)^{+0.48}_{-0.70}$ & 4.3 & --- & --- & --- & --- & --- \\
            \hline
            \end{tabular}
            }
            \caption{Summary of this work's main results. All the values reported here correspond to the data set combination:  Planck (TT,TE,EE+lowE+lensing), SPT-3G (TT+TE+EE), BAO (6dFGS+DR7+DR12+DR16), and SN Ia. For each of our models, we report in the second column the mean, best fit\tablefootnote{For the models for which we did not run an explicit minimization algorithm, we report the value of the best-fitting sample from the MCMC.} and 68\%CL credible interval of $H_0$. The third column presents the Marginalised Posterior Compatibility Level between SH0ES and the rest of the data, expressed in units of $\sigma$. The third column presents a frequentist version of the same test. The numbers in boldface show the cases reducing the tension below the $3\sigma$ level. The next columns show the $\chi^2$ and Akaike Information Criterion improvement for each model relative to $\Lambda$CDM, either without or with SH0ES data included. We refer the reader to Figure~\ref{fig:Graphical_Summary} for a graphical representation of the first four columns, and to section~\ref{Sec:Data and Method} for more details.}
            \label{Table:Summary}
    \end{table}

The first line of Table~\ref{Table:Summary} shows that the Hubble tension is now of the order of $6\sigma$ for $\Lambda$CDM with our baseline data set. Like in~\cite{H0_Olympics}, we will consider that a given model brings the tension down to an acceptable level when it decreases to $3\sigma$ or less. This condition may sound too permissive since a 3$\sigma$ tension is still significant. However, we need to set a conservative threshold given that in the future we may get new insight on unaccounted systematics in current data, that may not completely solve the tension, but bring it from an apparent 3$\sigma$ level to less than 2$\sigma$. Thus, to be on the safe side, we will not describe models with a 3$\sigma$ tension as clearly incompatible with all existing data. This threshold could be imposed to either of the $Q_\mathrm{MPCL}$ or $Q_\mathrm{DMAP}$ metrics, but we wish to remain agnostic concerning the use of a Bayesian versus frequentist metric. Thus, we will conservatively consider that our main test is passed whenever, for a given model, one of $Q_\mathrm{MPCL}$ or $Q_\mathrm{DMAP}$ is smaller or equal to $3\sigma$. This is the case only for three models in Table~\ref{Table:Summary}: $\Lambda$CDM+$m_e$+$\Omega_K$, $\Lambda$CDM+$m_e$+$\Omega_K$+$\Sigma m_\nu$ and EDE.

At this point, we can already conclude that none of the models within our list but these three remain compatible with all current data sets including SH0ES. More generally, we find that for all our models (including the ones still passing the test) the level of tension has grown since the analysis of ref.~\cite{H0_Olympics}. In addition, in the case of the Majoron model, more accurate theoretical predictions make the model less compatible with a high value of $H_0$. In ref.~\cite{H0_Olympics}, both the $\Lambda$CDM+$m_e$ and Majoron models did pass the 3$\sigma$-compatibility test. According to this work, they no longer appear as plausible solutions, with the Majoron results being in accordance with the findings of~\cite{Sandner:2023ptm}. The $\Lambda$CDM+$m_e$+$\Omega_K$ and EDE models barely pass our tests and with a worse tension level than in ref.~\cite{H0_Olympics}. 

We stress that the three surviving models do pass the 3$\sigma$-compatibility test, but only thanks to a larger error bar on $H_0$ rather than predicting a larger mean value for it. One can check this from the second column of Table~\ref{Table:Summary}, in which the mean $H_0$ value is only $68.2\,$km/s/Mpc for $\Lambda$CDM+$m_e$+$\Omega_K$ and EDE. The mean $H_0$ value is however larger for the $\Lambda$CDM+$m_e$+$\Omega_K$+$\Sigma m_\nu$ model, which predicts $H_0=69.75^{+1.85}_{-2.93}\,$km/s/Mpc.

Table~\ref{Table:Summary} also shows that $\Delta$AIC$_\mathrm{w/o}$ is always positive. This means that none of the models considered in this work is significantly preferred over $\Lambda$CDM in the absence of SH0ES data, even models that reduce the tension. This is another reason to consider these potential solutions as not very appealing: at present time, they appear as ad-hoc solutions, motivated only by the reduction of the Hubble tension, and not confirmed by any hint in other data sets. We may notice that the EDE model better fits our baseline data set than $\Lambda$CDM by $\Delta\chi^2=-4.6$, but given that this model has three additional free parameters, this is not statistically significant.

Finally, the 8$^\text{th}$ column of Table~\ref{Table:Summary} always returns large negative values of $\Delta$AIC$_\mathrm{w/}$, which  shows an improvement in goodness-of-fit compared to $\Lambda$CDM. However, as long as $\Delta$AIC$_\mathrm{w/o}$ remains small, we know that the improvement of the AIC$_\mathrm{w/}$ metric can only come from a reduction of the SH0ES $\chi^2$ value. For instance, when the tension is reduced from the 6$\sigma$ to the $n\sigma$ level, there is a potential to decrease the AIC$_\mathrm{w/}$ metric by about $6^2-n^2$. This is very consistent with the numbers appearing in the last column of Table~\ref{Table:Summary} for $\Delta$AIC$_\mathrm{w/}$. The fact that we do not obtain smaller numbers confirms that there is nothing in the data that the extended models can fit better than $\Lambda$CDM, except the SH0ES likelihood itself.

Our key conclusions from this work can be summarized by four points. First, we confirm the previous findings of ref.~\cite{H0_Olympics} that none of the \CE{}s of $\Lambda$CDM can reduce the Hubble tension to an acceptable level. Second, the $\Lambda$CDM+$m_e$ and Majoron models, which were strong candidates in ref.~\cite{H0_Olympics}, are no longer able to reduce the tension to an acceptable level. Third, the $\Lambda$CDM+$m_e+\Omega_\mathrm{K}$ and EDE models are still able to bring the tension down to the 2-3$\sigma$ level, but there is no hint in favor of these models in our combined Planck, SPT, BAO, and SNIa data set. Fourth, the $\Lambda$CDM+$m_e+\Omega_\mathrm{K}$+$\Sigma m_\nu$ model, which was never considered before, is the one in our list with the best potential to reduce the tension, with $Q_\mathrm{MPCL}=1.5\sigma$ and a mean value as large as $H_0=69.8\,$km/s/Mpc,  but like the previous models it does not improve the fit of non-SH0ES data compared to $\Lambda$CDM.
However, the three models $\Lambda$CDM+$m_e$+$\Omega_\mathrm{K}$, $\Lambda$CDM+$m_e$+$\Omega_\mathrm{K}$+$\Sigma m_{\nu}$ and EDE are still worth investigating in future works, both theoretically and with additional data. 

Before entering a more detailed discussion, we emphasize two points. First, although changes in the tension level compared to ref.~\cite{H0_Olympics} are mainly driven by the improved SH0ES measurements~\cite{New_H0}, the inclusion of SPT-3G 2018 and SDSS-DR16~\cite{SDSS_DR16} data plays a crucial role in our conclusions. For instance, compared to ref.~\cite{H0_Olympics}, the fact that the tension level exceeds the $3\sigma$ threshold for the $\Lambda$CDM+$m_e$ is due to the inclusion of SDSS-DR16 data.\footnote{For this model, one can check that, even if we use the same SH0ES likelihood as in ref.~\cite{H0_Olympics}, the $Q_\mathrm{MPCL}$ metric grows from 2.9 to 3.2 by just adding SPT-3G 2018 and eBOSS DR16 data.} On the other hand, as we explain in more detail in section~\ref{Sec:Results}, the improved polarization measurements from SPT-3G 2018 straighten the constraints on $m_e+\Omega_\mathrm{K}$ and EDE. 
Second, it is not the purpose of this paper to discuss the $S_8$ tension~\cite{S8_1,S8_2} in as many details as the Hubble tension, but the short discussion provided in section~\ref{sec:S_8_Tension} suggests that none of the models presented here exacerbates this other tension beyond the $3\sigma$ level.

\section{Cosmological models}
\label{Sec:Models}

     \begin{table}[hbt!]
        \centering
            \renewcommand{\arraystretch}{1.5}
            \begin{tabular} {l |c }
            \hline
            Parameter &  Priors \\
            \hline
            {$\Omega_bh^2$} &  $\mathcal{U}$ (0.005, 0.1)\\
            
            {$\Omega_ch^2$} & $\mathcal{U}$ (0.001, 0.99) \\
            
            {$\theta_s$} & $\mathcal{U}$ (0.5, 10) \\
            
            {$n_s$} & $\mathcal{U}$ (0.8, 1.2) \\
            
            {$\ln (10^{10}A_s)$} & $\mathcal{U}$ (1.67, 3.91) \\
            
            {$\tau_{\text{reio}}$} & $\mathcal{U}$ (0.01, 0.8) \\
            lowE $\tau_{\text{reio}}$ & $\mathcal{N}$ (0.054, 0.007)\\
             \hline
            \end{tabular}
            \caption{Priors on the cosmological parameters of our baseline $\Lambda$CDM model. For data sets that do not include Planck, we use a Gaussian prior on $\tau_{\text{reio}}$ specified in the last row.}
            \label{Priors_Cosmo_Params}
    \end{table}

We consider eleven extensions of the minimal $\Lambda$CDM model.
Our baseline $\Lambda$CDM model is parameterized with the usual six vanilla parameters,
\begin{equation}
            \mathcal{P}_{\text{cosmo}}=\big[\Omega_bh^2,\ \Omega_ch^2,\ \theta_s,\ n_s,\ A_s, \ \tau_{\text{reio}}\big],
            \label{Eq: Cosmo_params}
\end{equation}
 where $\Omega_b$ and $\Omega_c$ are the fractional densities of baryons and CDM, respectively, $h=H_0/$ (100km/s/Mpc) is the reduced Hubble parameter, $\theta_s$ is the angular scale of the sound horizon at decoupling (eq.~\eqref{Eq:Theta}), $n_s$ and $A_s$ are respectively the spectral index and amplitude of the primordial curvature power spectrum, and $\tau_{\text{reio}}$ is the photon optical depth to the reionization epoch. In this model, we assume three degenerate massive neutrino species with a total mass $\Sigma m_{\nu}=0.06\,$eV, close to the minimum allowed by oscillation experiments~\cite{Nu_Oscillation1,Nu_Oscillation2}. Our prior on each free parameter is described in Table~\ref{Priors_Cosmo_Params}. We denote a uniform prior between $a$ and $b$ as $\mathcal{U} (a, b)$ and a Gaussian prior centered on $a$ with standard deviation $\sigma$ as $\mathcal{N} (a,\sigma)$. By default, we adopt uniform priors on each parameter. Among the data sets considered in our analysis, Planck is the only one constraining $\tau_{\text{reio}}$, mainly through the measurement of the reionization bump in the low-multipole range of the polarization spectrum. Thus, whenever the Planck likelihood is not included in our data set, we adopt a Gaussian prior on $\tau_{\text{reio}}$ of the form $\mathcal{N}$ (0.054, 0.007) \cite{Planck_Likelihood}, referred to as the lowE $\tau$-prior.

We divide our eleven non-minimal models into two categories, the first being that of classical extensions of $\Lambda$CDM with compelling physical ingredients such as massive neutrinos, relativistic relics, dynamical dark energy, and spatial curvature. Previously, extra relativistic relics, spatial curvature, or dynamical dark energy with a Chevalier-Polarski-Linder  (CPL) equation of state~\cite{CPL1,CPL2} were suggested as possible solutions to the Hubble tension. However, it was later shown that each of these extensions (without involving additional ingredients) are too simple to solve this tension without raising other ones~\cite{H0_Olympics,CPL_No_1,CPL_NO_2,Planck2018}. Nevertheless, their free parameters are sufficiently well motivated to deserve an updated analysis, including the SPT-3G 2018 data. The results obtained with these models are also useful for comparison with those of more speculative models considered later on. In this first category, we study: 
\begin{enumerate}

\item $\Lambda$CDM$+\Sigma m_{\nu}$\\
Neutrino oscillation experiments~\cite{Nu_Oscillation1,Nu_Oscillation2} have established that neutrinos have a mass that can significantly affect cosmological observables~\cite{Bond:1980ha,Hu:1997mj,Nu_Julien_Sergio,Lesgourgues:2013sjj,Lattanzi:2017ubx}. Fixing $\Sigma m_{\nu}$ to $0.06\,$eV is based on the minimum value allowed by oscillation experiments. In this model  (and in many of the others considered later on), we consider $\Sigma m_{\nu}$ as a free parameter limited to positive values, while sticking to the assumption of neutrinos degenerate in mass. Note that the range $0<\Sigma m_{\nu}<0.06\,$eV is excluded by laboratory data, but is still interesting to consider in a cosmological analysis, given that we are not sure that all the assumptions of our baseline cosmological model are fulfilled. Note also that oscillation experiments imply that one should consider different masses for each of the three neutrino eigenstates, but adopting three equal masses has been proved to be an excellent approximation to the CMB and matter power spectra \cite{Lesgourgues:2004ps,Nu_Julien_Sergio,Lesgourgues:2013sjj,Archidiacono:2020dvx}.\\
\textit{Priors:} $\Sigma m_{\nu}\in \ \mathcal{U}$ (0, 1) eV.

\item $\Lambda$CDM$+\Sigma m_{\nu}+ w_0+w_a$.\\
 In this model, we additionally assume dynamical DE with a CPL~\cite{CPL1,CPL2} parametrization of the equation of state,
      \begin{equation}
          w (a) =w_0+w_a (1-a/a_0)~,
      \end{equation}
where $w_0$ and $w_a$ are free parameters and $a_0$ is the scale factor, $a$, today. Note that the effect of perturbations has been accounted for using the Parameterized Post-Friedmann formulation~\cite{Post_Friedmann}.\\
\textit{Priors:} $\Sigma m_{\nu}\in \ \mathcal{U}$ (0, 1) eV, $w_0\in \ \mathcal{U}$ (-2, -0.33), $w_a\in \ \mathcal{U}$ (-3, 2).
            
\item $\Lambda$CDM$+\Sigma m_{\nu}+\Omega_K$\\
On top of massive neutrinos, this model features spatial curvature with a parameter $\Omega_K$. Note that with a positive $\Omega_K$ -- corresponding to negative curvature -- one could, in principle, obtain the same angular diameter distance to the last scattering surface as with an increased value of $H_0$. \\
\textit{Priors:} $\Sigma m_{\nu}\in \ \mathcal{U} (0, 1)$ eV and $\Omega_K\in \ \mathcal{U}$ (-1, 1).
            
\item $\Lambda$CDM$+\Sigma m_{\nu}+N_\mathrm{eff}$\\  
Still, on top of massive neutrinos, this model assumes ultra-relativistic (ur) relics that are decoupled and free-streaming. Such relics are very easy to motivate since many extensions of the standard model of particle physics predict not only stable non-relativistic relics (dark matter) but also ultra-relativistic ones (dark radiation). Their contribution is accounted for as an increase in the effective neutrino number $N_\mathrm{eff}$ with respect to the standard model value $N_\mathrm{eff}=3.044$~\cite{N_eff1,N_eff2}. Since it is possible to increase simultaneously $H_0$ and $N_\mathrm{eff}$ without changing $\theta_s$ (see eq.~\eqref{Eq:Theta}) \cite{Hou:2011ec,Lesgourgues:2013sjj}, this model would be an obvious solution to the Hubble tension if high values of $N_\mathrm{eff}$ were not actually rising tension with the measurement of the damping tail in the CMB spectrum~\cite{Hou:2011ec,Lesgourgues:2013sjj,Planck2018,Trouble_H0,DiValentino_Neff}.\\
\textit{Priors:} $\Sigma m_{\nu}\in \ \mathcal{U} (0, 1)$ eV, $N_\mathrm{ur}\in \ \mathcal{U}$ (0, 2), where $N_\mathrm{ur}=N_\mathrm{eff}-3.044$.
            
\item $\Lambda$CDM$+\Sigma m_{\nu}+N_\mathrm{SIDR}$\\   
This model differs from the previous one through the assumption that the ultra-relativistic relics are strongly self-interacting~\cite{SIDR_ETHOS,SIDR}. Such relics are implemented in \texttt{CLASS} as an ``interacting dark radiation''  (IDR) species. At the level of the background, this model is identical to one with free-streaming relics and $N_\mathrm{eff}=3.044+N_\mathrm{SIDR}$, but perturbations undergo a slightly different evolution. Specifically, dark radiation clusters more when it is self-interacting, which slightly enhances the small-scale CMB spectrum; at the same time, neutrino over-densities propagate at a smaller velocity, and the neutrino drag effect is reduced~\cite{Bashinsky:2003tk,Lesgourgues:2013sjj,Audren:2014lsa,Baumann:2015rya,Follin:2015hya}. Both effects allow to fit CMB data better than with an equivalent amount of free-streaming relics, and thus to reach potentially higher values of $N_\mathrm{eff}$ and $H_0$.\\
\textit{Priors:} $\Sigma m_{\nu}\in \ \mathcal{U}$ (0, 1) eV and $N_\mathrm{SIDR}\in \ \mathcal{U}$ (0, 1.5).
\end{enumerate}  

Concerning the second category, we consider models that are more elaborate or include speculative physical ingredients, which recently received much attention for their ability to solve -- or at least reduce -- the Hubble tension. These models, which we call \EMs, all involve some dynamical mechanisms to reduce the comoving sounds horizon $r_s$ in eq.~\eqref{Eq:Theta}. Their ability to reduce the tension was already compared in~\cite{H0_Olympics}. Here, we wish to update this comparison in the presence of more recent CMB, BAO and SH0ES data sets. For some of the models, we will also complete the analysis by adding the summed neutrino mass to the list of free parameters. The main features of each model can be summarized as follows  (we refer the reader to~\cite{H0_Olympics} and references therein for further details): 

\begin{enumerate}
\setcounter{enumi}{5}
\item \textit{Varying} $m_e$\\
Several high-energy theories predict a possible variation of fundamental constants over time. For instance, in the framework of higher dimensional theories such as string theory, the appearance of light scalar fields that couple to other fundamental fields, such as leptons and quarks, is almost inevitable (see~\cite{VarMe1} and references therein for more details). These fields may take a long time to stabilise, and their dynamics can cause a variation in effective particle masses, particularly that of the electron. Without invoking extra dimensions, a drift in the electron mass $m_e$ can also appear in several implementations of the Higgs mechanism~\cite{Me_Higgs}. Since the gaps between energy levels in the hydrogen atom depend on $m_e$, a variation of this mass between the beginning of matter domination and today can shift the redshift of recombination and photon decoupling. If $m_e$ slowly decreases over time, recombination takes place at a larger redshift, and the sound horizon at decoupling is smaller. This can be compensated by an enhanced value of $H_0$ \cite{VarMe1,Planck_VarMe,VarMe2015,VarMe2018}. Since current cosmological data are insensitive to the detailed evolution of $m_e$ between recombination and today, it is customary to assume an approximately constant value of $m_e$ around recombination, another constant value today, and a step-like transition in between. In this case, the only new measurable parameter in this model is the ratio of the $m_e$ before, $m_\text{e,early}$, and after, $m_\text{e,late}$, this transition:  $m_\text{e,early}/m_\text{e,late}$. This model is implemented in \texttt{CLASS} with a transition taking place by default at redshift $z\sim 50$, but this arbitrary choice has no observable consequence (as long as it is after the end of recombination). Note that BBN also provides constraints on variations of the electron mass, with a roughly similar constraining power as CMB and BAO data (see for example \cite{VarMe_BBN}). Here we are not considering such additional constraints (assuming, for instance, that $m_e$ could also be different at BBN and CMB times). Note that a different electron mass at high redshift can change the time of electron-positron annihilation and thus the value of $\Delta N_\mathrm{eff}$ at later times. Using the \texttt{PRyMordial} code \cite{burns2023prymordial}, we estimate this shift to be of the order of $\Delta N_\mathrm{eff} / \Delta m_e \sim 0.1$. Our results will show that $m_e$ is allowed to vary typically by 2\% (when including at least BAO data), resulting in variations of $\Delta N_{\rm eff}$ of the order of 0.002. Since such small variations have a negligible impact on our observables, we can approximate $\Delta N_{\rm eff}$ as independent of $m_\text{e,early}/m_\text{e,late}$.\\
            \textit{Priors:} $m_\text{e,early}/m_\text{e,late} \in \ \mathcal{U}$ (0, 1.5).  
            
\item \textit{Varying} $m_e + \Sigma m_{\nu}$\\   
Since there is no strong reason to fix the total neutrino mass to $0.06\,\mathrm{eV}$, our next model features $\Sigma m_{\nu}$ as an additional free parameter. This offers an interesting opportunity to check for parameter degeneracies between $m_e$ and $\Sigma m_{\nu}$ given current data.\\
\textit{Priors:} m$_\text{e,early}/$m$_\text{e,late}\in \ \mathcal{U}$ (0.5, 1.5), $\Sigma m_{\nu}\in \ \mathcal{U}$ (0, 1) eV.

\item \textit{Varying} $m_e+\Omega_K$\\
  It was already noticed in~\cite{VarMe_Omk} that when fitting recent cosmological data, one finds a degeneracy between the ratio $m_\text{e,early}/m_\text{e,late}$ and the curvature parameter $\Omega_K$, which allows to reach even large values of $H_0$. Indeed, in the combined model, one can adjust both parameters, separately modifying the sound horizon and angular diameter distance to decoupling -- that is, with the numerator and denominator of eq.~\eqref{Eq:Theta}. We thus consider this combination of free parameters, first with a total neutrino mass fixed to $0.06\,\mathrm{eV}$.\\
    \textit{Priors:} $m_\text{e,early}/m_\text{e,late} \in \ \mathcal{U}$ (0.5, 1.5), $\Omega_K \in \ \mathcal{U}$ (-0.02,0.02)
            
\item \textit{Varying} $m_e+\Sigma m_{\nu}+\Omega_K$\\
Here we additionally vary the total neutrino mass since, as already mentioned, there is no strong theoretical motivation to stick to the minimum value of  $\Sigma m_{\nu}$. \\
 \textit{Priors:} $m_\text{e,early}/m_\text{e,late} \in \ \mathcal{U}$ (0, 1.5), $\Sigma m_{\nu}\in \ \mathcal{U}$ (0, 2) eV\footnote{We doubled the prior range on $\Sigma m_{\nu}$ for this model to make sure that we are not excluding allowed regions in parameter space.} $\Omega_K\in \ \mathcal{U}$ (-0.2, 0.2).

 \item \textit{Early Dark Energy}  (EDE)\\
In this case, a scalar field -- potentially motivated by string theory~\cite{AxiString} -- 
contributes to the expansion rate over a short period before recombination. This results in a reduction in the sound horizon that can be compensated by a higher value of $H_0$. In order to obtain temperature and polarization spectra compatible with CMB data, the scalar field must decay quickly at some critical time $t_\mathrm{c}$ parameterized by the scale factor $a_\mathrm{c}=a (t_\mathrm{c})$  (assuming that $a (t_0)=1$). EDE models were first proposed as solutions to the Hubble tension in ref.~\cite{EDE3} and have been extensively studied since then, see e.g.~\cite{EDE1,NEDE,Elisa-Eiishiro,Elisa-Laura,Birefrengence1,Birefrengence2,EDE_Curvature}. We stick to the initial model of ref.~\cite{EDE3}, parameterized by the initial value of the scalar field $\theta_{\mathrm{i}}$, the scale factor at which the scalar field starts to decay $a_{\mathrm{c}}$, and the fractional energy of EDE at this time $f_{\text{EDE}} (a_{\mathrm{c}})$. We note that the EDE model studied in this work represents a large class of EDE models, but not all of them (see~\cite{Ups_Downs_EDE} for a review). A full investigation of other EDE models is beyond the scope of the current work.\\
            \textit{Priors:} $\theta_{\mathrm{i}}\in \ \mathcal{U}$ (0.01, 3.1), $\log (a_{\mathrm{c}})\in \ \mathcal{U}$ (-4.5, -3), $f_{\text{EDE}} (a_{\mathrm{c}})\in \ \mathcal{U}$ (0, 0.5).
            
 \item \textit{Majoron} \\
A possible scenario for the generation of neutrino masses invokes the spontaneous symmetry breaking of the global $U (1)$ lepton number symmetry, which produces a pseudo-Nambu-Goldstone scalar $\phi$ called the Majoron. The authors of refs.~\cite{Majoron1,Majoron2,Majoron3} investigated a version of this scenario compatible, as a byproduct, with enhanced values of $H_0$, with the Majoron mass $m_{\phi}\sim$ eV. A study of the various interactions shows that in this case, the population of Majorons gets efficiently produced at a temperature $T\sim m_\phi$ and decays into neutrinos soon afterward. This model provides a concrete implementation of a particle physics model in which the radiation density -- or the effective parameter $N_\mathrm{eff}$ -- gets enhanced over a short period before recombination, similar to the EDE model. Moreover, during this period, neutrinos behave as a self-interacting fluid, that is, like IDR, which may further help to increase $N_\mathrm{eff}$ and raise $H_0$. This effect is parameterized with an independent and dimensionless effective decay width, $\Gamma_\mathrm{eff}$. To be more specific, in this model, $N_\mathrm{eff}$ gets inherently enhanced by the energy density stored, firstly, in the Majoron particles when they thermalize, and later on, in the extra active neutrinos produced by their decay. This enhancement (combined with the effect of Majoron-induced neutrino scattering) helps to reduce the Hubble tension but is not sufficient to solve it.
To reach sufficiently high values of $N_\mathrm{eff}$, refs.~\cite{Majoron1,Majoron2} consider such a Majoron in combination with another population of stable relativistic relics that we call here Additional Dark Radiation (ADR). The constant contribution of ADR to $N_{\rm eff}$, dubbed $N_\mathrm{ADR}$, is the third free parameter specific to this model. 

In summary, the model assumes that $N_\mathrm{eff}$ starts from a value 3.044+$N_\mathrm{ADR}$ well before photon decoupling and experiences some further growth at a later time due to neutrino/Majoron conversions. The maximum value reached by $N_\mathrm{eff}$ depends on the three parameters $N_\mathrm{ADR}$, $m_\phi$ and $\Gamma_\mathrm{eff}$, and the $\Lambda$CDM model is recovered in the limit in which these three parameters go to zero.
To obtain a more natural scenario, the authors of ref.~\cite{Majoron3} noticed that the ADR component could account for an initial population of the same Majoron particles,\footnote{In numerical simulations, this amounts to initializing the density of Majoron at high redshift to a positive $N_\mathrm{ADR}$ instead of zero. In ref.~\cite{Majoron3}, our $N_\mathrm{ADR}$ is denoted as $\Delta N_\mathrm{eff}^\mathrm{BBN}$.} produced in the very early universe through a mechanism motivated by other ingredients in the low-scale Majoron model of leptogenesis and neutrino mass generation. In this case, the model only involves one new type of cosmological relics. This scenario was implemented in the Boltzmann solver \texttt{CLASS} by the authors of~\cite{Majoron1,Majoron2,Majoron3}. Recently, ref.~\cite{Sandner:2023ptm} updated this modified branch of \texttt{CLASS}, with a more accurate numerical treatment and a generalisation to the alternative scenario. Here, we stick to the Majoron case, but we use the updated code kindly provided by the authors of~\cite{Sandner:2023ptm}.
As a final remark on this model, we recall that the Majoron was historically introduced as a possible mechanism to generate neutrino masses. 
However, the main motivation behind the work of refs.~\cite{Majoron1,Majoron2,Majoron3,Sandner:2023ptm} was to study a model with non-trivial dynamics in the background and perturbation equations of dark radiation and neutrinos -- arising, in the present case, from the late-time out-of-equilibrium thermalization of a light boson interacting with neutrinos. While neutrino masses can be naturally incorporated within this framework, doing so can significantly complicate both the phenomenology and the computational overheads.
Here, we will stick to the assumption of negligible neutrino masses for this particular model.\\
\textit{Priors:} $\log (m_\mathrm{\phi}/\mathrm{eV})\in \ \mathcal{U}$ (-1, 1), $\log (\Gamma_\mathrm{eff})\in \ \mathcal{U}$ (-2, 2), and $N_\mathrm{ADR}\in \ \mathcal{U}$ (0, 1.5).
\end{enumerate}
Many additional interesting models have been proposed to reduce the Hubble tension. These include Early Modified Gravity~\cite{EMG}, New Early Dark Energy~\cite{NEDE}, other types of EDE~\cite{Recent_EDE_SPT,Vivian_Tristan_EDE,Hot_EDE},
Stepped Dark Radiation \cite{Aloni:2021eaq,Nils_WZDR},
Interacting Dark Energy \cite{Zhai:2023yny}, Primordial Magnetic Fields~\cite{Karsten1,Karsten2} and many others. Due to finite computational resources, we could not include these models in this analysis, but we will investigate them in future works.

\section{Methodology}
\label{Sec:Data and Method}

\subsection{Observational data}
\label{subsec:data}

We consider the following cosmological data sets:
\begin{enumerate}
    \item SPT: CMB power spectra from SPT-3G 2018 for TT, TE, and EE~\cite{SPT-3G:2022,SPT-3G:2021}.
    \item Planck: CMB power spectra from Planck 2018 for TT, TE, EE +lowE+lensing~\cite{Planck_Likelihood,Planck_Lensing}.
    \item ACT: CMB power spectrum data from the Atacama Cosmology Telescope DR4 for TT, TE, and EE~\cite{ACT1,ACT2}.
    \item BAO: Baryon Acoustic Oscillation data from 6dFGS~\cite{SDSS_6dF}, BOSS DR7~\cite{SDSS_DR7} and DR12~\cite{SDSS_DR12}, and eBOSS  DR16~\cite{SDSS_DR16}. We do not include redshift-space distortions~\cite{RSD} or Alcock-Paczynski~\cite{Alcock-Paczynski} measurements from this data set.
    \item Pantheon: SN Ia luminosity data from the Pantheon sample~\cite{Pantheon}.
    \item SH0ES: we also use the most recent estimate of $H_0$ by the SH0ES collaboration  ($H_0 = 73.29 \pm 0.90 $km/s/Mpc~\cite{New_H0}) for the purpose of estimating the tension with other data sets, but we do not combine it with other data sets when computing mean values and confidence intervals.
\end{enumerate}

Concerning a potential correlation between the different CMB experiments, we first note that the SPT-3G 2018 footprint is very small ($f_\mathrm{sky}=0.04$) compared to Planck, which is almost full sky. Therefore, correlations between these two data sets are negligible, as shown in~\cite{SPT-3G:2021}. On the other hand, there is no overlap in the footprints of  SPT-3G 2018 and ACT-DR4. There is instead an overlap between Planck and ACT footprints. The ACT collaboration recommends using their TT data only for $\ell>1800$ when combining with Planck. Given that this overlap is known to have only a small impact on parameter constraints, we neglect it for simplicity. To make the notation more concise, we define the following acronyms for the different data combinations: 
\begin{align}
    & \mathcal{D}_{\text{S}} = \text{SPT}; \nonumber\\
    & \mathcal{D}_{\text{SP}} = \text{SPT+Planck};\nonumber\\
    & \mathcal{D}_{\text{SB}} = \text{SPT+BAO};\nonumber\\
    & \mathcal{D}_{\text{PB}} = \text{Planck+BAO};\nonumber\\
    & \mathcal{D}_{\text{SPB}} = \text{SPT+Planck+BAO};\nonumber\\
    & \mathcal{D}_{\text{SPBP}} = \text{SPT+Planck+BAO+Pantheon};\nonumber\\
    & \mathcal{D}_{\text{SPAB}} = \text{SPT+Planck+ACT+BAO};\nonumber\\
    & \mathcal{D}_{\text{SPABP}} = \text{SPT+Planck+ACT+BAO+Pantheon}.
\label{Eq:Data_sets_name}
\end{align}

\subsection{Bayesian analysis pipeline}

To make theoretical predictions and compute observables, we rely on the Boltzmann codes \texttt{CAMB}\footnote{\url{https://github.com/cmbant/CAMB}}~\cite{CAMB}, \texttt{CLASS}\footnote{\url{https://github.com/lesgourg/class_public}}~\cite{CLASS1,CLASS2}, and a couple of modified branches of \texttt{CLASS}. For the $\Lambda$CDM model and its extensions with the parameters $\Sigma m_\nu$, $w_0$, $w_a$, $N_\mathrm{eff}$ and $\Omega_k$, one can in principle choose between the public versions of~\texttt{CAMB} and \texttt{CLASS}, which provide highly consistent results. We use \texttt{CAMB} \texttt{v1.3.6} for the $\Lambda$CDM+$w_0$+$w_a$ and $\Lambda$CDM+$\Sigma m_{\nu}$+$\Omega_K$ models, and \texttt{CLASS} \texttt{v3.2.1} in other cases. For extensions including at least the parameters $N_\mathrm{SIDR}$ or $m_e$, we used the public (\texttt{master}) branch of \texttt{CLASS} \texttt{v3.2.1}. For EDE, we used the \texttt{master} branch of  \texttt{AxiCLASS} \texttt{v3.2.1}\footnote{\url{https://github.com/PoulinV/AxiCLASS}}~\cite{AxiCLASS1,AxiCLASS2}, and for the Majoron model, a modified private branch of \texttt{CLASS} kindly provided by the authors of \cite{Sandner:2023ptm}.

We run our MCMCs using \texttt{COBAYA}~\cite{COBAYA1,COBAYA3} and its Metropolis-Hastings algorithm~\cite{Metropilis_Hastings1,Metropolis_Hastings2}. We use \texttt{GetDist}~\cite{GetDist} to calculate Bayesian posteriors, means, and confidence intervals. Our runs are usually based on 8 MCMCs, with a minimal convergence level given by the Gelman-Rubin criterion~\cite{Gelman_Rubin} $R-1\leq0.02$. Moreover, our results for the $\mathcal{D}_{\text{SPBP}}$ data set are derived from those for the $\mathcal{D}_{\text{SPB}}$ data set using importance sampling with the Pantheon likelihood. The outcome is trustworthy since, as we will see in section~\ref{Sec:Results}, the addition of the Pantheon data does not shift the means and errors by a significant amount. 

In addition to computing the Bayesian posteriors conventionally, we investigate the possibility of speeding up this computation by using a novel lightweight emulator introduced and described in ref.~\cite{günther2023uncertaintyaware}. Unlike other emulators that are used in CMB cosmology \cite{Spurio_Mancini_2022, G_nther_2022, Nygaard_2023, bonici2023capsejl}, which require a prior generation of training data and training of neural networks,\footnote{\cite{Nygaard_2023} uses an iterative sampling and training algorithm that converges on the relevant parameter space of the posterior distribution.} this implementation uses an online learning algorithm that samples training data from the MCMC itself. This enables the emulator to be accurate precisely in the region where the posterior is not negligible while keeping the training set small (just $\mathcal{O}(10^2)-\mathcal{O}(10^3)$ realisations in the investigated models). This is different from other inference algorithms such as those of refs.~\cite{christoph_weniger_2023_8349112, Karamanis_2021}, since our emulator predicts cosmological observables like $C_\ell$'s rather than posteriors. Hence, this emulator remains agnostic about the used sampling algorithm and the likelihoods. Thus, it benefits from the (relatively) small dimensionality of model parameter space compared to nuisance parameter space. Furthermore, the architecture of the emulator, which is based on Gaussian Processes~\cite{GP_Book}, allows for an estimate of the uncertainty of the prediction. The estimate can be incorporated into the sampling algorithm so that whenever the emulator's prediction is insufficiently accurate, new training data is generated. This procedure ensures very reliable performance for the emulator. The current code implementation is publicly available within a modified \texttt{COBAYA} repository.\footnote{\hyperlink{github.com/svenguenther/cobaya}{github.com/svenguenther/cobaya}} A future release will present a more modular version of the same emulator that will be compatible with \texttt{COBAYA}, \texttt{MontePython}~\cite{MontePython1,MontePython2} and additional samplers~\cite{guenther2024}. Like in ref.~\cite{günther2023uncertaintyaware}, we find that this emulator achieves a major speed-up of the inference process while delivering a highly accurate posterior estimate. In appendix \ref{APP: Pipeline_Plus}, we provide details on the performance and accuracy of the emulator. We also explain its benefits for performing assisted $\chi^2$ minimization in order to compute some of our tension metrics.

\subsection{Quantifying the tension level}
\label{sec:metrics}

There exist various statistical tools to assess the ability of a model to fit discrepant data sets. Some of these tools try to quantify the level of tension between different data assuming a given cosmological model, while others try to compare the goodness of fit of different models to a given data set, which could include discrepant data. Moreover, some estimators are more inspired by the Bayesian framework, and others by the frequentist approach. Here, for practical reasons, we limit ourselves to a non-exhaustive list of tests. We focus on a few metrics that are relatively easy to compute and match or generalise the ones used in ref.~\cite{H0_Olympics}, to allow for a straightforward comparison.


Using $\mathcal{D}_\mathrm{SPBP}$ -- defined in~\eqref{Eq:Data_sets_name} -- as our reference data set, we start with two metrics that assess the level of tension between SH0ES and non-SH0ES data for each model. Note, however, that models for which these metrics return good results are not necessarily appealing solutions to the Hubble tension, because they don't necessarily predict a high $H_0$ value. Nevertheless, such models are still compatible with all existing data.


\begin{enumerate}
    \item
{\it Marginalised Posterior Compatibility Level (MPCL)}.
We start discussing our first metric in a general context. Consider a parameter $x$ with two independent marginalised posteriors, ${\cal P}_1(x)$ and ${\cal P}_2(x)$, resulting from Bayesian inference with two different experiments, but for the same theoretical model. Second, consider independent realisations $x$ and $x'$ of the two posteriors, respectively, and compute the probability distribution ${\cal P}(\delta)$ of their difference $\delta=x'-x$. Mathematically, this distribution is given by
\begin{equation}
    {\cal P}(\delta) = {\cal N} \int dx \, {\cal P}_1(x)\,
{\cal P}_2(x-\delta),
\label{Eq:Convolution}
\end{equation}
where ${\cal N}$ is a normalization factor. A standard way to quantify the level of compatibility between the two posteriors ${\cal P}_1$ and ${\cal P}_2$ is to address the following question:\footnote{In particular, the estimator quoted later in eq.~\eqref{Eq:GT} and sometimes called the Gaussian Tension estimator finds its justification in this question; it provides the exact answer to it when the two posteriors are Gaussian.} According to the distribution ${\cal P}(\delta)$, what is the probability that $\delta$ belongs to the minimum credible interval that encompasses $\delta=0$? If ${\cal P}(\delta)$ peaks in the $\delta>0$ range, this interval reads $[0, \delta']$, otherwise it would be $[\delta',0]$ -- where, in both cases, $\delta'$ satisfies ${\cal P}(\delta')={\cal P}(0)$. Then, the probability $q$ addressing the previous question is given by:
\begin{equation}
    q=\int_0^{\mathrm{\delta'}}d\delta\, {\cal P}(\delta)~.
    \label{Eq:q-value}
\end{equation}
 Given this probability $q\in [0,1]$, one can conclude that, in the framework of the assumed model, the two measurements are compatible at the $100(1-q)\%$ level. For a more intuitive interpretation, this probability can be converted into a number $n$ of $\sigma$'s using\footnote{Strictly speaking, this conversion from probabilities to $\sigma$'s refers to Gaussian distributions, but in a wider context, this definition of $n$ can be adopted as a convention to represent a given probability $q$, even for non-Gaussian distributions.} 
\begin{equation}
    n=\sqrt{2}\,\mathrm{erf}^{-1}(q),
    \label{Eq:Generalized_GT}
\end{equation} 
where erf is the error function. Note that this metric was also used in the work of~\cite{Raveri_Doux,leizerovich2023tensions}. As an example, if $q=0.9973$, then the two measurements would be compatible at the 0.27\% level, corresponding to $n=3$. In fact, for the special case where ${\cal P}_1(x)$ and ${\cal P}_2(x)$ are both Gaussian, the result can be expressed as:\footnote{One way to see this is to recall that the difference between two normally distributed variables also obeys a normal distribution, with a mean and a standard deviation given by the numerator and denominator of eq.~\eqref{Eq:GT}, respectively.}  
    \begin{equation}
        n = \frac{\bar{x}_2 -\bar{x}_1}{\sqrt{\sigma_{x_1}^2+\sigma_{x_2}^2}} \,,
        \label{Eq:GT}
    \end{equation}
where $\bar{x}_i$ and $\sigma_{x_i}$ are respectively the mean and standard deviation of ${\cal P}_i(x)$.
The r.h.s of eq.~\eqref{Eq:GT} was called the Gaussian Tension (GT) metric (or $\Delta_{\text{GT}}$) in ref.~\cite{H0_Olympics}. In that work, this number was quoted for each model, with a warning that $\Delta_{\text{GT}}$ is a meaningful estimate of the compatibility between the two posteriors only in the Gaussian case. This is a severe limitation since our extended models tend to have strongly non-Gaussian marginalised posteriors for $H_0$. Here, we adopt the exact version of this metric by making use of a simple script that evaluates explicitly equations (\ref{Eq:Convolution}, \ref{Eq:q-value}). This provides the exact answer to the question formulated above, even when the posteriors are not Gaussian.

In the case considered here, $x$ stands for the Hubble parameter $H_0$. We thus set ${\cal P}_2(x)={\cal P}_\mathrm{SH0ES}(H_0)$, with the latter being the SH0ES likelihood, i.e. a Gaussian distribution with mean $H_{0,\mathrm{SH0ES}}=73.29$ km/s/Mpc and standard deviation $\sigma_\mathrm{SH0ES}=0.9$ km/s/Mpc. Moreover, ${\cal P}_1(x) ={\cal P}_\mathrm{SPBP}(H_0)$ is now the posterior distribution of $H_0$ estimated from one of our MCMC runs for a given model and for the data set $\mathcal{D}_\mathrm{SPBP}$. Since SH0ES and $\mathcal{D}_\mathrm{SPBP}$ are independent data sets, our previous definitions apply. Then, eq.~\eqref{Eq:Convolution} becomes 
\begin{equation}
{\cal P}(\delta) = {\cal N} \int dH_0 \, {\cal P}_\mathrm{SPBP}(H_0)\,
{\cal P}_\mathrm{SH0ES}(H_0-\delta)
\simeq {\cal N}' \sum_i w_i \, {\cal P}_\mathrm{SH0ES}(H_{0,i}-\delta)
\end{equation}
where we approximated the integral with a sum over each point $i$ in our MCMCs with weight $w_i$ and Hubble parameter $H_{0,i}$. The normalisation factor 
${\cal N}$ (${\cal N}'$) is trivial to compute by integrating over the whole range (summing over all chain points). Once the probability ${\cal P}(\delta)$ has been computed numerically, our script finds $q$ using eq.~\eqref{Eq:q-value}, and then converts it into a number $n$ of $\sigma$'s using eq.~\eqref{Eq:Generalized_GT} for the purpose of comparing this metric with $\Delta_\mathrm{GT}$. We call the final result Q$_\mathrm{MPCL}\equiv n$. We checked explicitly that for cosmological models with a Gaussian marginalised posterior on $H_0$, such as $\Lambda$CDM, we recover Q$_\mathrm{MPCL}=\Delta_{\text{GT}}$ as expected.

\item
{\it Difference of the Maximum A Posteriori (DMAP).}
As a second attempt to quantify the tension between SH0ES and other data for each model, we consider the ``difference of the maximum a posteriori'', $Q_{\text{DMAP}}$~\cite{Q_DMAP,H0_Olympics}. This number is given by the square root of the difference between the effective best-fit $\chi^2$ obtained for a given model and data set with and without a SH0ES prior. We perform this test for the $\mathcal{D}_\text{SPBP}$ data set, which amounts to defining
\begin{equation}
    Q_{\text{DMAP, model}} \equiv \sqrt{\chi^2_{\text{min,~model,~}\mathcal{D}_{\text{SPBP}+\text{SH0ES}}} - \chi^2_{\text{min,~model,~}\mathcal{D}_{\text{SPBP}}}}~.
    \label{Eq: Q_DMAP}
\end{equation}
This number quantifies how much the best fit worsens in the presence of SH0ES data and, thus, how much SH0ES is in tension with other data sets when assuming a model. 
For a given model, $Q_{\text{DMAP}}$ would ideally vanish if the fit to the data set ${\cal D}_{\rm SPBP}$ would return a best-fit value of $H_0$ equal to the SH0ES central value.

In the case where the marginalized $H_0$ posteriors are exactly Gaussian, one can prove mathematically that $Q_{\text{DMAP}}=Q_{\text{MPCL}}=\Delta_{\text{GT}}$. In other cases, the MPCL and DMAP metrics address the same question under a more Bayesian and frequentist point of view, respectively. Indeed, the MPCL depends on priors and volume effects (through the marginalisation over other parameters), while the DMAP uses only restricted information at the best fit. Thus, one can argue that each of the two metrics has its pros and cons, and it is useful to report and discuss both of them. However, due to the computational cost of minimizations in high dimensional spaces, we
report the DMAP metric solely for the more promising models. 

\item
{\it Akaike Information
Criterion (AIC).} With $Q_{\text{MPCL,model}}$ and/or $Q_{\text{DMAP, model}}$, one can compare the ability of different models to alleviate the tension. However, these tests do not allow us to compare the intrinsic goodness-of-fit of these models and do not penalize the presence of additional free parameters. The latter aspect is crucial because one can always decrease the best-fit $\chi^2$ by adding an arbitrary number of parameters, which would amount to over-fitting statistical fluctuations in the data.

To address these caveats, one can use additional metrics. 
On the frequentist side, a possibility is to use the Akaike Information Criterion (AIC)~\cite{AIC1,Bayesian2},
\begin{align}
    \Delta\text{AIC} = & \chi^2_{\text{min, model,~}\mathcal{D}} - \chi^2_{\text{min,~}\Lambda\text{CDM}, \mathcal{D}} \nonumber \\ 
    & + 2\big (N_{\text{model}} 
    -N_{\Lambda\text{CDM}}\big)~,
    \label{Eq:AIC}
\end{align}
defined for a given model and data set $\mathcal{D}$. This test quantifies the goodness-of-fit of a given model compared to $\Lambda$CDM. Additionally, it implements Occam's razor by penalizing models with a number of parameters $N_{\text{model}}$ bigger than that of $\Lambda$CDM, $N_{\Lambda\text{CDM}}$, thus complementing the previous two metrics.
We compute the AIC for two data sets, $\mathcal{D}=\mathcal{D}_{\text{SPBP}}$ and $\mathcal{D}=\mathcal{D}_{\text{SPBP+SH0ES}}$, in order to address separately two interesting questions: would our data set prefer a given extended model independently of SH0ES data? And does it prefer this model in the presence of SH0ES data? However, once more, we compute the AIC only for the most promising models, to reduce the number of minimizations. For brevity, we denote the AIC metric with a subscript ``w/'' (``w/o'') when the SH0ES likelihood is included (omitted) in the minimization, i.e. $\Delta$AIC$_\mathrm{w/}$ ($\Delta$AIC$_\mathrm{w/o}$).
\end{enumerate}

The four criteria mentioned above (MPCL, DMAP, AIC$_\mathrm{w/}$ and AIC$_\mathrm{w/o}$) already give a reasonable assessment of how a given model is performing in terms of fitting the data and reducing the Hubble tension. However, if a model reduces the Hubble tension by having a larger error bar on $H_0$, compared to $\Lambda$CDM, without actually predicting a larger value, we cannot objectively say that it is solving the Hubble tension, nor that it is preferred over $\Lambda$CDM \cite{Cortes:2023dij}. Rather, the only conclusion we can make is that this model is not excluded and that more data is required to properly assess it.

It is important to emphasize that the list of criteria presented in this section is neither complete nor conclusive. In particular, the quantity $-2\ln\mathcal{L}$ loses its traditional interpretation in terms of $\chi^2$-statistics in the presence of highly non-Gaussian posteriors. Nevertheless, taken together, these tests provide a fairly good indication of how well the considered models match the currently available data. 

\section{Results and Discussion}
\label{Sec:Results}

\begin{figure}[hbt!]
        \centering
        \includegraphics[width=15cm,angle=0]{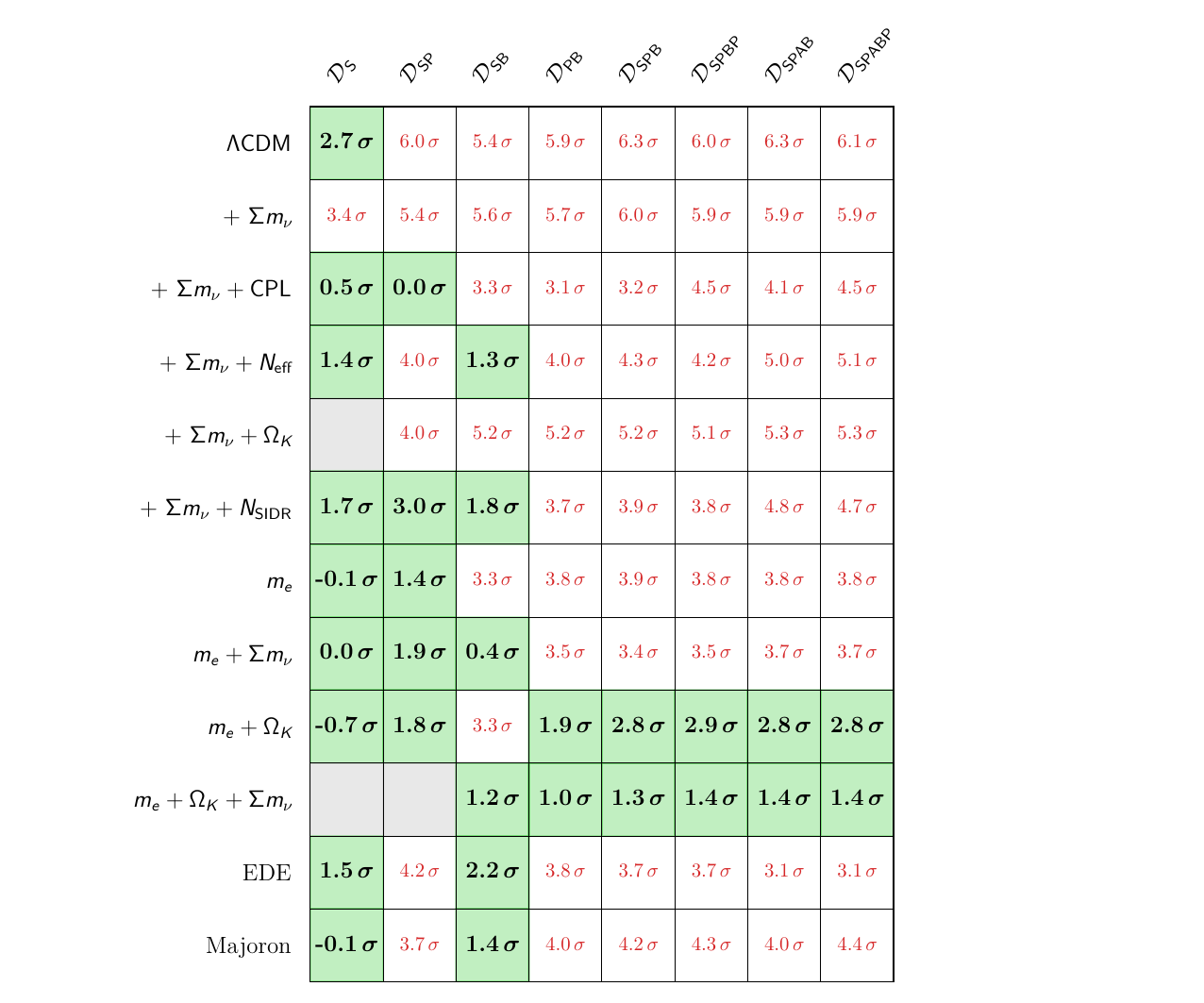}
        \caption{Tension between the $H_0$ measurement by the SH0ES~\cite{New_H0} collaboration and the marginalised $H_0$ posterior obtained in this work for each of the considered models and data combinations. The tension is quantified by $Q_\text{MPCL}$, see eq.~\eqref{Eq:GT}, and is given in units of $\sigma$. We highlight in bold the cases where the tension level is smaller or equal to 3$\sigma$. Note that empty entries are cases we did not run due to their high computational cost. By convention, we add a minus sign in the few cases for which the posterior peaks at a value larger than the SH0ES value.}
        \label{fig:Graph_Table3}
\end{figure}

\begin{figure}[ht!]
        \centering
        \includegraphics[width=13cm,angle=0]{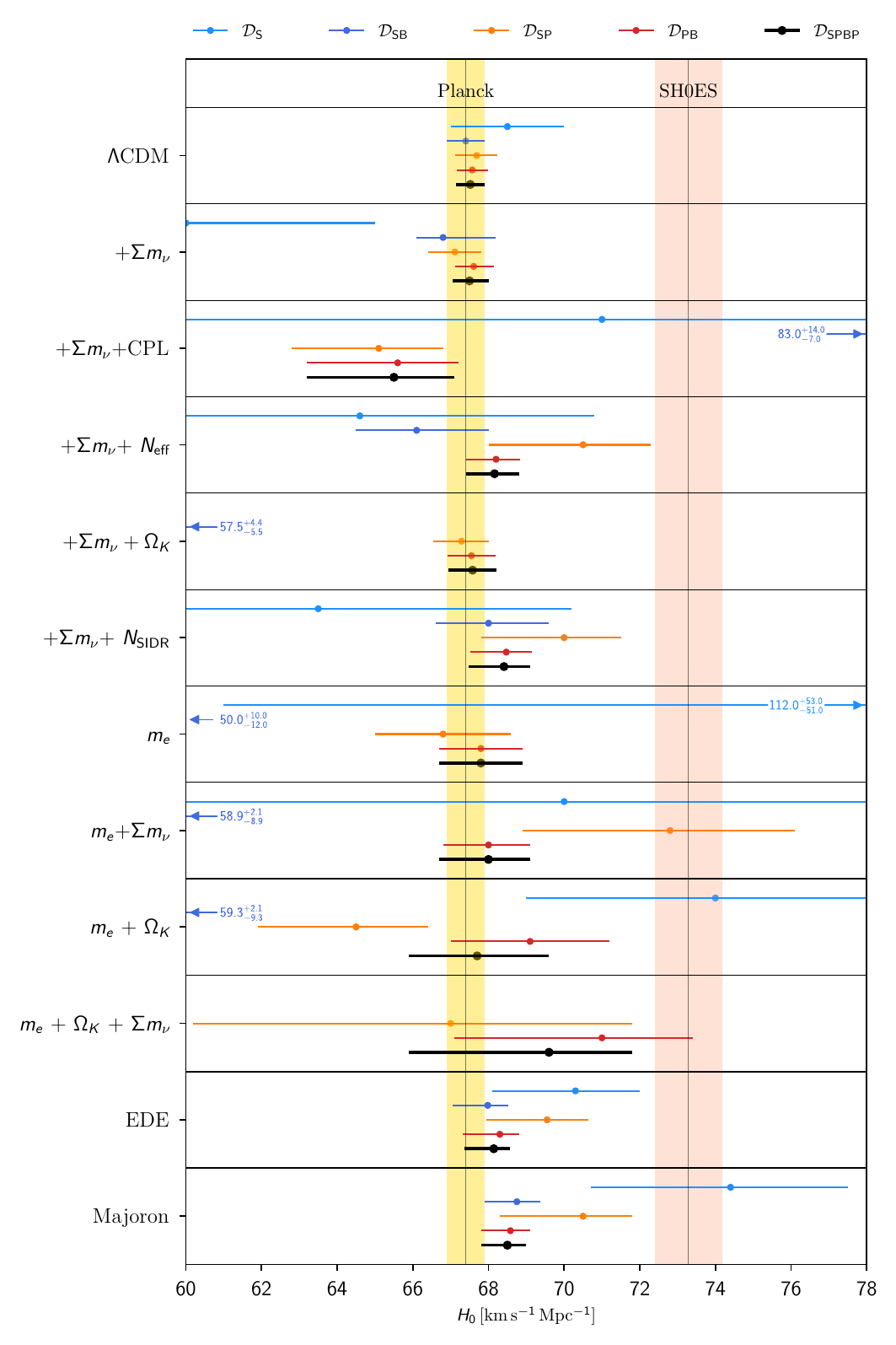}
        \caption{Graphical representation of the first four columns of Table~\ref{Table:H0_values}, as well as our baseline data set $\mathcal{D}_\mathrm{SPBP}$. In particular, we highlight the difference in constraining power between Planck and SPT-3G 2018 for each studied model. 
        \label{fig:Graph_Table4}}
\end{figure}

In this section, we elaborate on our main findings, starting with a presentation in Figure~\ref{fig:Graph_Table3} of the MPCL tension metric of $H_0$ for each model and data combination. 
The DMAP and $\Delta$AIC metrics were already reported in Table~\ref{Table:Summary} (with a graphical summary in Figure~\ref{fig:Graphical_Summary}), but only for a selection of interesting cases and for the $\mathcal{D}_\mathrm{SPBP}$ data set. Specifically, Table~\ref{Table:Summary} includes additional metrics only for the models with $Q_\mathrm{MPCL}<4\sigma$, which are the SIDR model, all models with a varying $m_e$, and the EDE model.
 Finally, we present the mean value and credible interval for $H_0$ for each model and data combination in Table~\ref{Table:H0_values}, with a graphical summary in Figure~\ref{fig:Graph_Table4}. 
 

We would like to emphasize that our work is not only relevant for assessing the tension level of various extended cosmological models given recent CMB, BAO, and SH0ES data but also for updating the constraints on the parameters of these models. For this purpose, we provide all bounds on additional parameters (beyond the six $\Lambda$CDM ones) derived from the data set $\mathcal{D}_\mathrm{SPBP}$ in Table~\ref{Table:Additional_Params}.

%
%
%
%
%

     \begin{table}[hbt!]
        \resizebox{\textwidth}{!}{%
            \centering
            \renewcommand{\arraystretch}{1.5}
            \begin{tabular} {l |c c c c c >{\bf}c c c}
            \hline
            Models &  $\mathcal{D}_{\text{S}}$ & $\mathcal{D}_{\text{SP}}$ & $\mathcal{D}_{\text{SB}}$ & $\mathcal{D}_{\text{PB}}$ & $\mathcal{D}_{\text{SPB}}$ & $\mathcal{D}_{\text{SPBP}}$& $\mathcal{D}_{\text{SPAB}}$ & $\mathcal{D}_{\text{SPABP}}$\\
            \hline
            {$\Lambda$CDM} & $68.5^{+1.5}_{-1.5}$ & $67.40^{+0.49}_{-0.50}$ & $67.69^{+0.55}_{-0.56}$ & $67.57^{+0.41}_{-0.41}$ & $67.52^{+0.37}_{-0.37}$ & 67.56$^\mathbf{+0.35}_\mathbf{-0.38}$ & $67.49^{+0.34}_{-0.39}$ & $67.53^{+0.34}_{-0.37}$\\
            {+$\Sigma m_{\nu}$} & $60.0^{+5.0}_{-5.6}$ & $66.8^{+1.4}_{-0.7}$ & $67.11^{+0.71}_{-0.70}$ & $67.61^{+0.53}_{-0.48}$ & $67.50^{+0.52}_{-0.44}$ & 67.60$^\mathbf{+0.49}_\mathbf{-0.43}$ & $67.50^{+0.58}_{-0.44}$ & $67.59^{+0.53}_{-0.42}$\\
            {+$\Sigma m_{\nu}$+CPL} & $71^{+10}_{-15}$ & $83^{+14}_{-7}$ & $65.1^{+1.7}_{-2.3}$ & $65.6^{+1.6}_{-2.4}$ & $65.6^{+1.6}_{-2.4}$ & 67.94$^\mathbf{+0.78}_\mathbf{-0.79}$ & $66.5^{+1.3}_{-1.7}$ & $67.92^{+0.81}_{-0.81}$\\
            {+$\Sigma m_{\nu} +$ $N_{\text{eff}}$} & $64.6^{+6.2}_{-7.0}$ & $66.1^{+1.9}_{-1.6}$ & $70.5^{+1.8}_{-2.5}$ & $68.20^{+0.63}_{-0.78}$ & $68.16^{+0.65}_{-0.76}$ & 68.25$^\mathbf{+0.62}_\mathbf{-0.76}$ & $67.83^{+0.58}_{-0.60}$ & $67.93^{+0.57}_{-0.58}$\\ 
            {+$\Sigma m_{\nu}+ \Omega_{\text{k}}$} & ---& $57.4^{+4.4}_{-5.5}$ & $67.29^{+0.73}_{-0.74}$ & $67.55^{+0.63}_{-0.63}$ & $67.58^{+0.64}_{-0.64}$ & 67.67$^\mathbf{+0.62}_\mathbf{-0.62}$ & $67.59^{+0.64}_{-0.64}$ & $67.69^{+0.62}_{-0.62}$ \\
            {+$\Sigma m_{\nu}$+ $N_{\text{SIDR}}$} & $63.5^{+6.7}_{-6.8}$ & $68.0^{+1.6}_{-1.4}$ & $70.0^{+1.5}_{-2.2}$ & $68.47^{+0.68}_{-0.95}$ & $68.41^{+0.70}_{-0.93}$ & 68.53$^\mathbf{+0.69}_\mathbf{-0.92}$ & $67.86^{+0.60}_{-0.61}$ & $67.96^{+0.57}_{-0.58}$\\
            {$m_e$} & 112$^{+53}_{-51}$ &
            $50^{+10}_{-13}$ & $66.8^{+1.8}_{-1.8}$ & $67.8^{+1.1}_{-1.1}$ & $67.8^{+1.1}_{-1.1}$ & 68.0$^\mathbf{+1.1}_\mathbf{-1.1}$ & $67.7^{+1.1}_{-1.1}$ & $67.9^{+1.1}_{-1.1}$\\
            {$m_e$+$\Sigma m_{\nu}$} & $70^{+20}_{-20}$ & $58.9^{+2.1}_{-8.9}$ & $72.8^{+3.3}_{-3.9}$ & $68.0^{+1.1}_{-1.2}$ & $68.0^{+1.1}_{-1.3}$ & 68.2$^\mathbf{+1.1}_\mathbf{-1.2}$ & $68.0^{+1.2}_{-1.2}$ & $68.2^{+1.2}_{-1.2}$\\
            {$m_e$ + $\Omega_k$} &
            $74^{+16}_{-5}$ & $59.3^{+2.1}_{-9.3}$ & $64.5^{+1.9}_{-2.6}$ & $69.1^{+2.1}_{-2.1}$ & $67.7^{+1.9}_{-1.8}$ & 68.2$^\mathbf{+1.6}_\mathbf{-1.6}$ & $67.5^{+1.9}_{-1.9}$ & $68.1^{+1.6}_{-1.6}$\\
            {$m_e$ + $\Omega_k$ + $\Sigma m_{\nu}$} & --- & --- & $67.0^{+4.8}_{-6.8}$ & $71.0^{+2.4}_{-3.9}$ & $69.6^{+2.2}_{-3.7}$ & 69.8$^\mathbf{+1.8}_\mathbf{ -2.9}$ & $69.5^{+2.3}_{-3.7}$ & $69.8^{+2.0}_{-3.0}$\\ 
            {EDE} & $70.3^{+1.7}_{-2.2}$ & $67.98^{+0.54}_{-0.92}$ & 69.6$^{+0.9}_{-1.6}$ & $68.3^{+0.52}_{-0.98}$ & $68.12^{+0.43}_{-0.78}$ & 68.18$^\mathbf{+0.42}_\mathbf{ -0.79}$ & $68.7^{+0.6}_{-1.4}$ & $68.8^{+0.6}_{-1.4}$ \\
            {Majoron} & $74.4^{+3.1}_{-3.7}$ & $68.75^{+0.62}_{-0.86}$ & $70.5^{+1.3}_{-2.2}$ & $68.58^{+0.53}_{-0.77}$ & $68.50^{+0.48}_{-0.70}$ & 68.55$^\mathbf{+0.48}_\mathbf{-0.70}$ & $68.6^{+0.46}_{-0.64}$ & $68.64^{+0.48}_{-0.61}$\\
            \hline
            \end{tabular}
            }
            \caption{Mean values of $H_0$ in km/s/Mpc for each model and each combination of data set along with its standard deviation.\label{Table:H0_values}}
    \end{table}
    
We also show in figures~\ref{fig:Mnu_Nsidr_Main} to \ref{fig:maj_sup_main} the 2D confidence contours on $H_0$, S$_8$ and the additional parameters of each model for the data set combinations $\mathcal{D}_\mathrm{PB}, \ \mathcal{D}_\mathrm{SPB}, \ \mathcal{D}_\mathrm{SPBP}$ and $\mathcal{D}_\mathrm{SPABP}$.\footnote{We focus on these because they are the most relevant to draw conclusions about the performance of each model.} Further plots and results are provided in the appendices and supplementary \texttt{github} repository~\cite{Sup_Plots_Github}.

As a general comment, we should highlight that none of the models considered in this work are preferred over $\Lambda$CDM, at least as long as SH0ES data is not taken into account. Thus, the confidence interval on each extended parameter is always compatible with the value this parameter would have in the minimal $\Lambda$CDM model -- 
that is, 
a negligible neutrino mass, a vanishing spatial curvature, a plain cosmological constant, no extra relativistic relics, a constant electron mass, etc. 

\begin{table}[hbt!]
        \resizebox{\textwidth}{!}{%
            \centering
            \renewcommand{\arraystretch}{1.5}
            \begin{tabular} {l |l }
            \hline
            Models & Additional Parameters\\
            \hline
            {$\Lambda$CDM}& ---\\
            {+$\Sigma m_{\nu}$} & $\Sigma m_{\nu} < 0.16$ eV (95\%) \\
            {+$\Sigma m_{\nu}$+CPL}& $\Sigma m_{\nu} < 0.29$ eV (95\%), $w_0 = -0.97 \pm 0.08$, $w_a = -0.29 \pm 0.39$ \\
            {+$\Sigma m_{\nu} +$ $N_{\text{eff}}$} & $\Sigma m_{\nu}  < 0.15$ eV (95\%) , N$_{\text{eff}} < 0.17\ (95\%) $\\
            {+$\Sigma m_{\nu}$+ N$_{\text{SIDR}}$} & $\Sigma m_{\nu} < 0.15\ \mathrm{eV} (95\%)$, $N_{\text{SIDR}} < 0.16 \ (95\%)$ \\
            {+$\Sigma m_{\nu}+ \Omega_{K}$} & $\Sigma m_{\nu} < 0.17$ eV (95\%), $\Omega_\mathrm{K} = -0.0005 \pm 0.0020$ \\
            {$m_e$} & $\mathrm{m}_\mathrm{e,early}/\mathrm{m}_\mathrm{e,late} = 1.003 \pm 0.006$ \\
            {$m_e$+$\Sigma m_{\nu}$} & $\mathrm{m}_\mathrm{e,early}/\mathrm{m}_\mathrm{e,late} = 1.0057 \pm 0.0090$,  $\Sigma m_{\nu} < 0.29 \ \mathrm{eV} (95\%)$\\
            {$m_e$ + $\Omega_K$} & $\mathrm{m}_\mathrm{e,early}/\mathrm{m}_\mathrm{e,late} = 1.0035 \pm 0.0164 $,  $\Omega_\mathrm{K} = -0.0005 \pm 0.0048 $\\
            {$m_e$ + $\Omega_K$ + $\Sigma m_{\nu}$} & $\mathrm{m}_\mathrm{e,early}/\mathrm{m}_\mathrm{e,late} = 1.03 \pm 0.03$, $\Omega_\mathrm{K} = -0.004 \pm 0.006$, $\Sigma m_{\nu} < 0.48$ eV (95\%)\\
            {EDE} & $\theta_{\mathrm{i}} = 1.8 \pm 0.9$, $\log (a_{\mathrm{c}}) = -3.8 \pm 0.4 $, $f_{\text{EDE}} (a_{\mathrm{c}}) < 0.06$ (95\%)  \\
            {Majoron} & $\log(m_\mathrm{\phi}/\mathrm{eV})=0.2950 \pm 0.6598$, $\log(\Gamma_\mathrm{eff})=0.0556 \pm 0.8846$, $\Delta N_\mathrm{ADR} < 0.15 \ (95\%)$  \\
            \hline
            \end{tabular}}
            \caption{Updated constraints on additional parameters for each model considered in this work. We present the mean and standard deviation of each parameter, except for $\Sigma m_{\nu}$ and $f_{\text{EDE}} (a_{\mathrm{c}})$, for which we present the 95\% CL upper bound. These constraints are from the data set combination $\mathcal{D}_{\text{SPBP}}$. }
            \label{Table:Additional_Params}
\end{table}

\subsection{General trends\label{sec:generalities}}

The level of the Hubble tension has grown significantly over the last couple of years. For $\Lambda$CDM, we find that even without Planck data but with the SPT+BAO combination ${\cal D}_{SB}$, the tension is as large as $Q_\mathrm{MPCL}=5.4\sigma$. For all combinations including at least Planck, $Q_\mathrm{MPCL}$ fluctuates between $5.9\sigma$ and $6.3\sigma$. Since fitting $\Lambda$CDM to the data leads to nearly Gaussian posteriors, the two metrics quantifying the tension give approximately the same result, with $Q_\mathrm{MPCL}=6.0\sigma$ versus $Q_\mathrm{DMAP}=5.8\sigma$ for the ${\cal D}_\mathrm{SPBP}$ data set. For the more restricted data set ${\cal D}_\mathrm{PB}$, which was already considered in Table~2 of reference~\cite{H0_Olympics}, the tension has grown from $Q_\mathrm{DMAP}=4.1\sigma$ in that reference to $Q_\mathrm{DMAP}=5.8\sigma$ in the present work. The only difference between the two analyses is the inclusion of additional BAO data from eBOSS DR16 and the new SH0ES measurement. We checked that the increase in the tension metrics is driven by the latter.

In the next sections, we discuss the tension metrics for each of our extensions of $\Lambda$CDM. As a foreword, we stress a few general trends that can be immediately inferred from Table~\ref{Table:Summary}. Here, we limit ourselves to the discussion of the ${\cal D}_\mathrm{SPBP}$ data set. Since $\Lambda$CDM has a $Q_\mathrm{DMAP}=5.8\sigma$ tension with such data, we know that a model solving the tension could in principle reduce the best-fit $\chi^2$ by about $5.8^2 \simeq 34$. In the column reporting the $\Delta \chi^2_\mathrm{w/o}$ of each model without SH0ES, we see that the extended models considered here provide a very small or even negligible improvement to the minimum $\chi^2$. In other words, they do not provide a significantly better fit to the ${\cal D}_\mathrm{SPBP}$ data set, and they are not preferred over $\Lambda$CDM when the SH0ES measurement is disregarded, since they all have a positive $\Delta $AIC$_\mathrm{w/o}$. This also implies that the $\Delta \chi^2_\mathrm{w/}$ computed with SH0ES data only reflects the reduction of the tension with SH0ES, and is not a better fit to the other data sets. Thus, we can roughly expect $\Delta \chi^2_\mathrm{w/}$ for a given model to be of the same order of magnitude as $Q_{\mathrm{DMAP,\, }\Lambda\mathrm{CDM}}^2-Q_\mathrm{DMAP,\,model}^2$. One can check that this relation is indeed approximately fulfilled in each line of Table~\ref{Table:Summary}. 

In particular, a model reducing the tension from the $\Lambda$CDM level to a marginally acceptable level of $\sim 3\sigma$ will have a $\Delta \chi^2_\mathrm{w/}$ of the order of $3^2-5.8^2\simeq 25$. Then, as long as this model has a reasonable number of parameters, it will have a large negative $\Delta$AIC$_\mathrm{w/}$. We should not over-interpret this result as a proof that extended models are clearly preferred. It merely reflects the fact that the tension is so large that any model leading to larger error bars for $H_0$ will reduce the minimum $\chi^2$ by many units and will easily pass the $\Delta$AIC test. We should thus put more weight on the other metrics, which tell us how compatible a model is with the full data set, and whether there would be indications in favor of this model even without SH0ES data.

\subsection{Classical extensions of $\Lambda$CDM}

The first outcome of our work is a confirmation of the findings of ref.~\cite{H0_Olympics} that simple extensions of $\Lambda$CDM, which combine neutrinos of arbitrary mass with CPL dynamical dark energy, extra relativistic relics or spatial curvature, do not ease the Hubble tension. 

As a general trend, SPT or SPT+BAO data remain compatible with large $H_0$ values in models with massive neutrinos plus dynamical dark energy or extra relativistic relics. With SPT+BAO data, the mean value of $H_0$ is even larger than $70\,$km/s/Mpc in the $\Lambda$CDM+$\Sigma m_\nu$+$N_{\rm eff}$ or $N_{\rm SIDR}$ models (see Table~\ref{Table:H0_values}, column ${\cal D}_{\rm SB}$). However, the combination of SPT+BAO data contains limited information on the angular scale of the first and second acoustic peak of the temperature auto-correlation power spectrum .\footnote{The complicated modeling of correlated atmospheric noise across frequencies required removing a lot of the signal on those scales. See Section II of~\cite{SPT-3G:2022} for more details.} When full-sky Planck data are added, the strong constraining power of these scales brings back the confidence intervals of $H_0$ to lower values. 

For the category of models considered here, adding SPT data to Planck+BAO data only marginally affects the results, and actually increases the tension, since the mean values and error bars on $H_0$ become systematically smaller (see the first six lines in Figure~\ref{fig:Graph_Table3} and Table~\ref{Table:H0_values} and compare the columns ${\cal D}_{\rm PB}$ and ${\cal D}_{\rm SPB}$). Finally, adding ACT data exacerbates the tension slightly. Thus, we can say that while Planck+BAO data alone already disfavor large values of $H_0$ within \CE{}s of $\Lambda$CDM, the better measurement of the CMB damping tail by SPT and ACT confirms this trend.

As a result, all \CE{} models have  Q$_\mathrm{MPCL}\geq 3.2\sigma$ with SPT+Planck+BAO data,  Q$_\mathrm{MPCL}\geq 3.8\sigma$ when adding Pantheon data, or  Q$_\mathrm{MPCL}\geq 4.5\sigma$ with the final addition of ACT data. We can also see in Table~\ref{Table:H0_values} that with all data combinations that include at least Planck+BAO, the mean value of $H_0$ always remains in the range from 65.5 to 68.5~km/s/Mpc. 

More specifically, taking into account the effect of neutrinos with arbitrary mass in a flat or non-flat cosmology never allows to raise the mean value of $H_0$ significantly above the one predicted by $\Lambda$CDM. The mean value even tends to be smaller when $\Sigma m_\nu$ is let free. As a matter of fact, increasing the neutrino mass while fixing the quantities best probed by CMB anisotropy data (angular scale of the sound horizon, redshift of equality, and baryon-to-photon ratio) requires a smaller value of $H_0$. Thus, when fitting most cosmological models with massive neutrinos to data, one usually observes a negative correlation between $\Sigma m_\nu$ and $H_0$. As soon as BAO data is included, the MPCL is no better with the $\Lambda$CDM+$\Sigma m_\nu$(+$\Omega_k$) models than in the minimal $\Lambda$CDM case.

On the other hand, when combining the effect of neutrinos with arbitrary mass with CPL dark energy, we confirm previous results from~\cite{CPL_No_1}, that is: CMB data alone is compatible with a wide degeneracy between $\Sigma m_\nu$, DE parameters and $H_0$, and predicts surprisingly large values of $H_0$, even when the new SPT data is included (with $\mathcal{D}_\mathrm{SP}$ we get $H_0=83.3^{+13.9}_{-7.3}$~km/s/Mpc). But as soon as BAO data is also included, this degeneracy is lifted and the MPCL metric is high again (for instance,  Q$_\mathrm{MPCL}= 4.5\sigma$ for $\mathcal{D}_\mathrm{SPBP}$).

\begin{figure}[hbt!]
        \centering
        \includegraphics[width=12cm,angle=0]{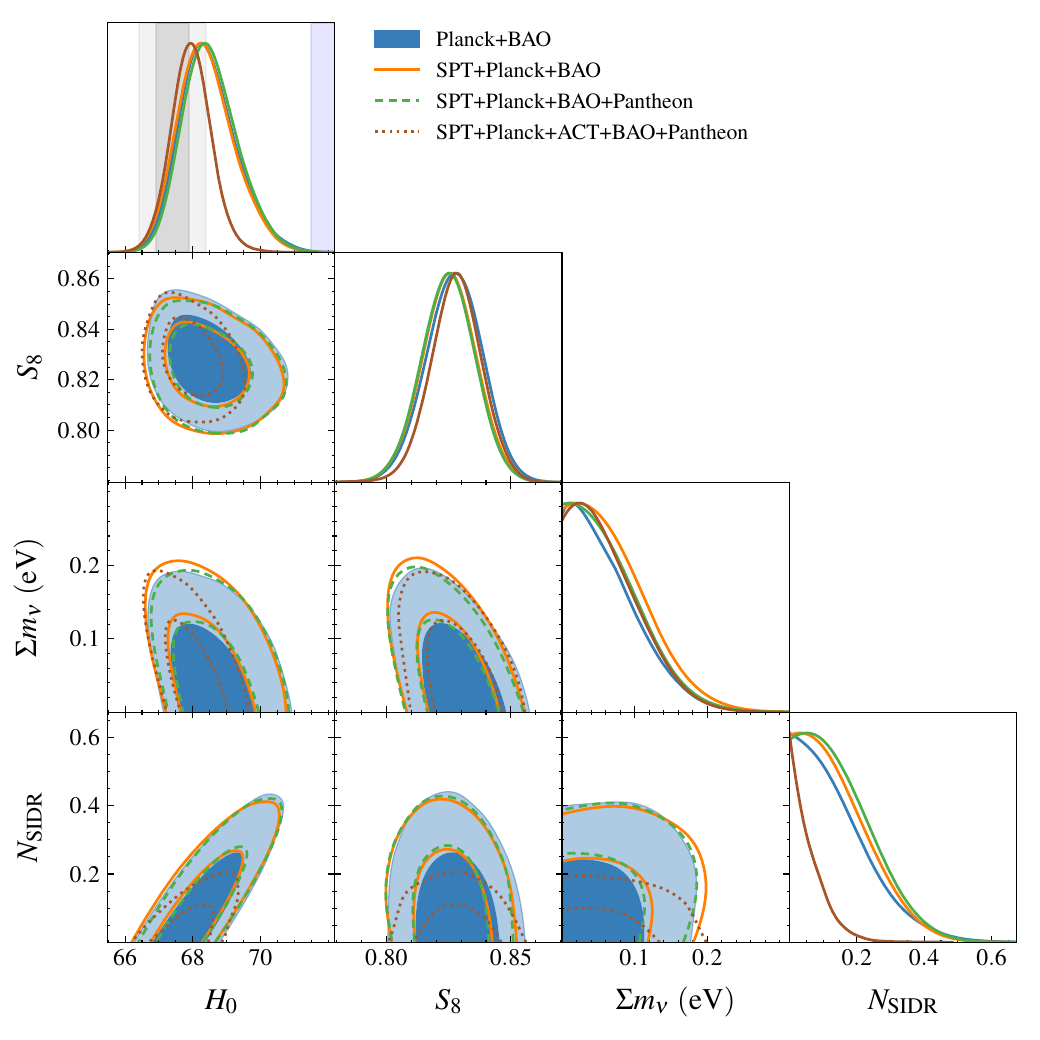}
        \caption{$\Lambda$CDM+$\Sigma m_{\nu}$+$N_{\rm SIDR}$: marginalised posteriors and 68\%/95\% CL contours on $H_0$, $\mathrm{S}_8$, and the two additional parameters of this model, $N_\mathrm{SIDR}$ and $\Sigma m_{\nu}$.  In the top left panel ($H_0$ marginalised posterior), the grey bands corresponds to the 68\%/95\% confidence intervals on $H_0$ from Planck~\cite{Planck2018}, while the purple ones correspond to those of SH0ES~\cite{New_H0}.}
        \label{fig:Mnu_Nsidr_Main}
\end{figure}

Finally, when combining the effect of neutrinos with arbitrary mass and free-streaming ($N_\mathrm{eff}$) or self-interacting ($N_\mathrm{SIDR}$) Dark Radiation (DR), we confirm that when at least the Planck data is taken into account, a degeneracy opens up between the DR density parameter and $H_0$. As expected, this degeneracy is even stronger when the DR is self-interacting. We find that the addition of recent SPT data preserves this degeneracy (see Figure~\ref{fig:Mnu_Nsidr_Main}). For instance, in the model with SIDR and with SPT+Planck, we get $H_0 = (68.0^{+1.6}_{-1.4})$ km/s/Mpc and a reduction of the tension down to $Q_\mathrm{MPCL}=3.0\sigma$. However, the degeneracy is partially lifted by adding BAO data, such that $Q_\mathrm{MPCL}$ increases to $3.9\sigma$, and totally lifted when further adding ACT data  ($Q_\mathrm{MPCL}=4.8\sigma$): unlike SPT data, ACT measures a CMB damping tail incompatible with a large DR density (for a study on the compatibility between ACT and SPT, see~\cite{SPT-3G:2022}). Focusing on the data set $\mathcal{D}_\mathrm{SPBP}$, we find that for the SIDR model, the Hubble tension remains large according to both metrics, with $Q_\mathrm{MPCL}=3.8\sigma$ and $Q_\mathrm{DMAP}=4.0\sigma$. This model is doing better than $\Lambda$CDM when the SH0ES likelihood is included, with $\Delta \mathrm{AIC}_\mathrm{w/}=-13.1$, but given the high value of the other metrics it cannot be seen as a solution to the tension. Thus, Dark Radiation alone (or in combination with massive neutrinos) cannot solve the Hubble tension, unless additional ingredients are invoked. These could be, for instance, DR isocurvature modes like in~\cite{SIDR_Review}, or additional couplings in the dark sector such as the Majoron~\cite{Majoron1,Majoron2,Majoron3} or the stepped Dark Radiation~\cite{Buen-Abad:2022kgf,Buen-Abad:2023uva,Schoneberg:2023rnx} models. However, these scenarios fall in the category of \EMs, which we now delve into.

\subsection{Varying $m_e$ and its extensions}
  \label{SubSec:Me_Results}

\begin{figure}[hbt!]
                \centering
                \includegraphics[width=12cm,angle=0]{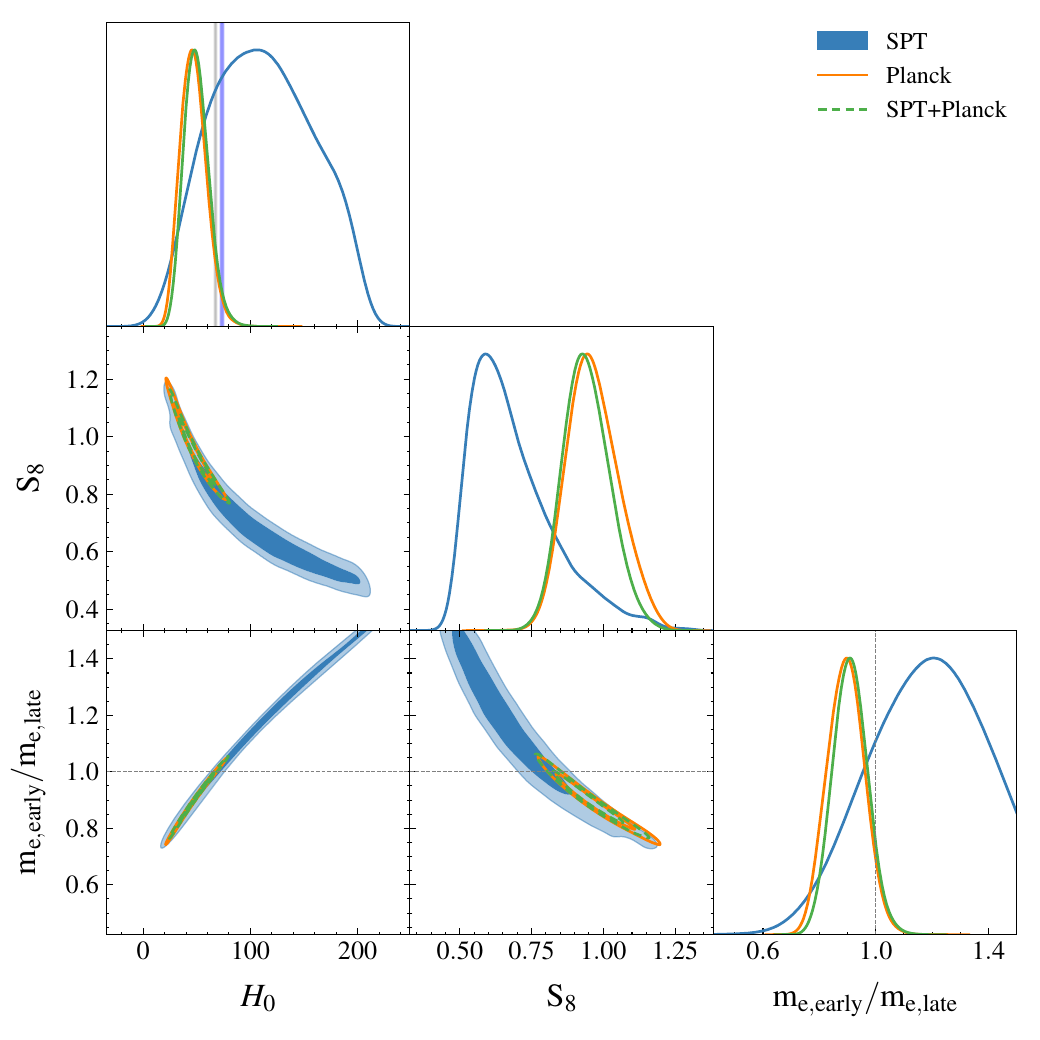}
                \caption{Same as Figure \ref{fig:Mnu_Nsidr_Main} for the varying $m_e$ model and only Planck and SPT data.}
                \label{fig:Me_SPTvsPlanck_Main}
\end{figure}


\begin{figure}[hbt!]
        \centering
        \includegraphics[width=12cm,angle=0]{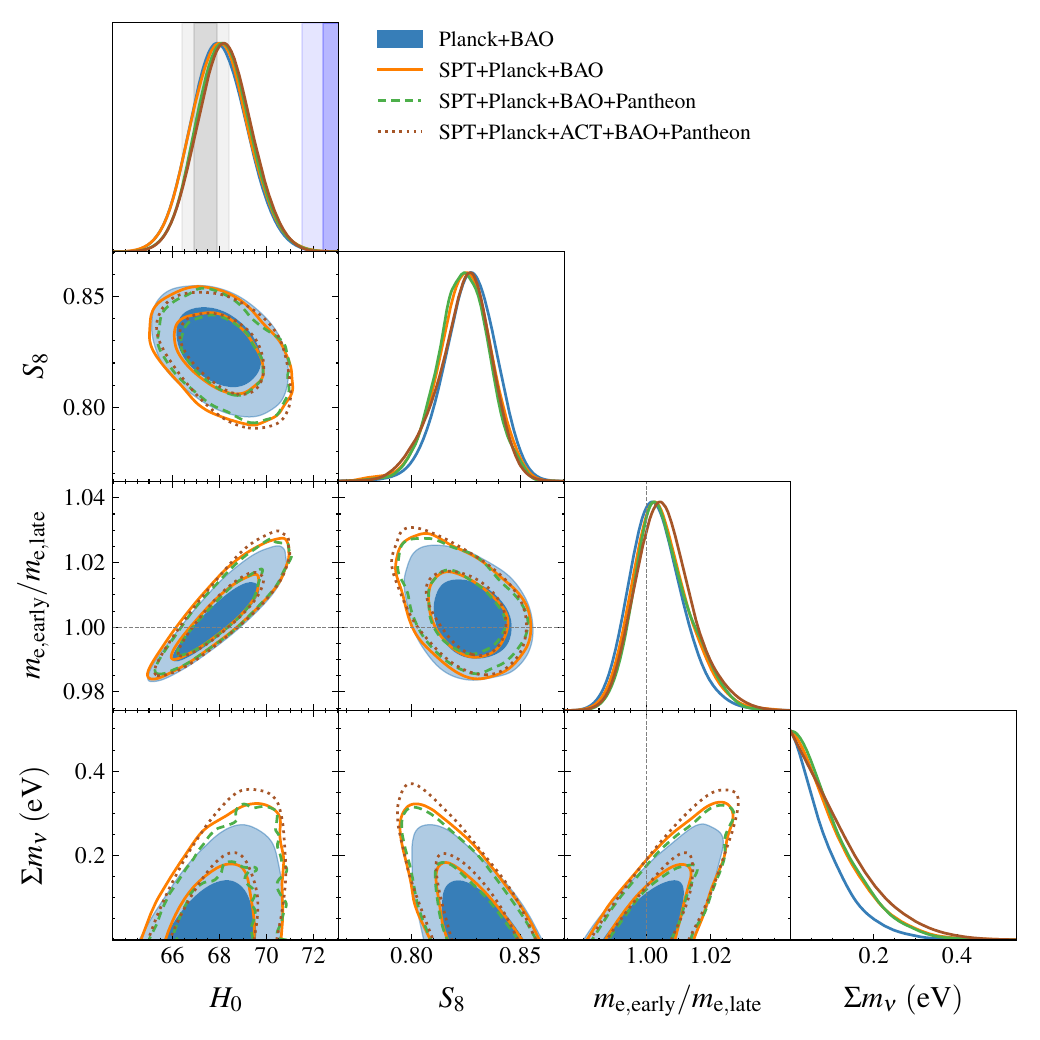}
        \caption{Same as Figure \ref{fig:Mnu_Nsidr_Main} for the varying $m_e$+$\Sigma m_{\nu}$ model.}
        \label{fig:Me_Mnu_Main}
\end{figure}

\begin{figure}[hbt!]
                \centering
                \includegraphics[width=12cm,angle=0]{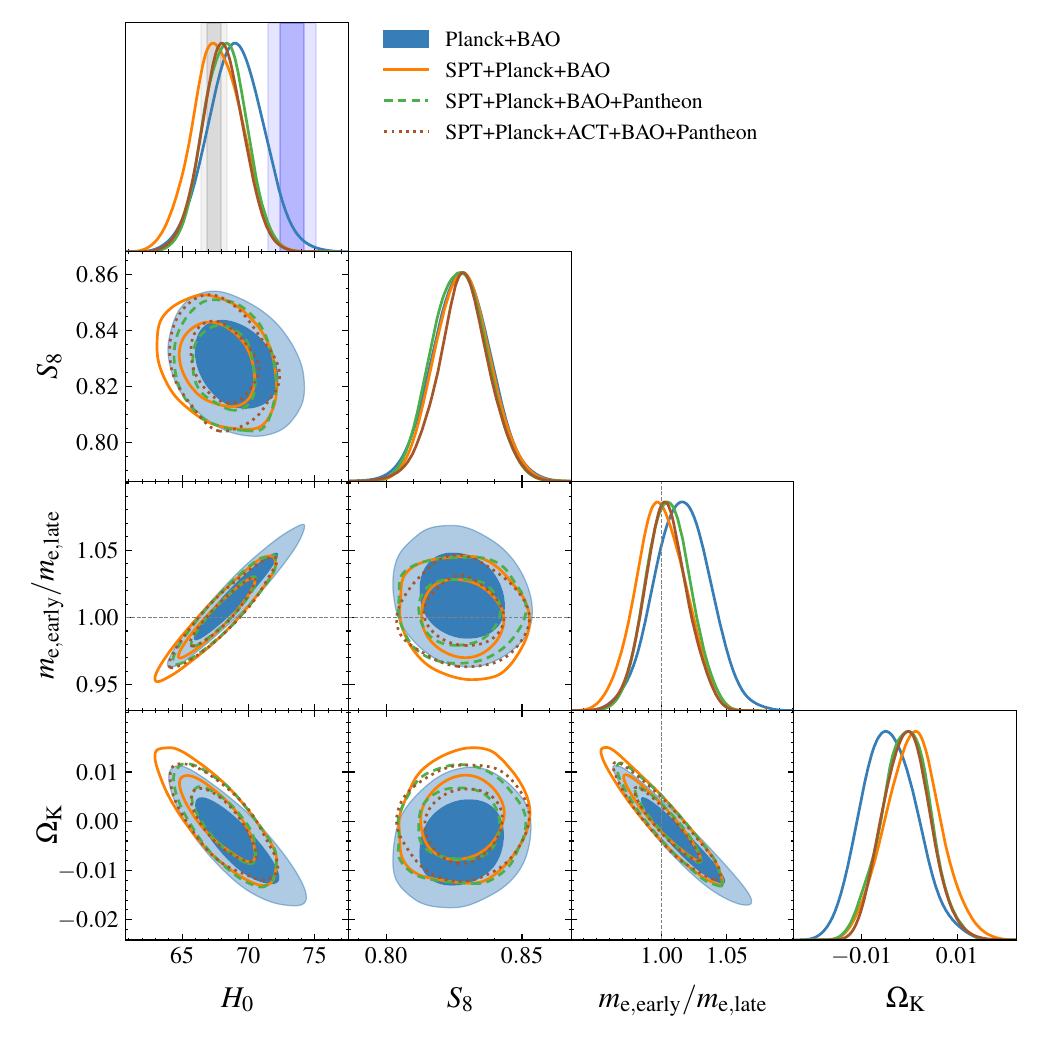}
                \caption{Same as Figure \ref{fig:Mnu_Nsidr_Main} for the varying $m_\mathrm{e} + \Omega_k$ model.}
                \label{fig:Me_Omegak_main}
            \end{figure}

\begin{figure}[hbt!]
               \centering
               \includegraphics[width=15cm,angle=0]{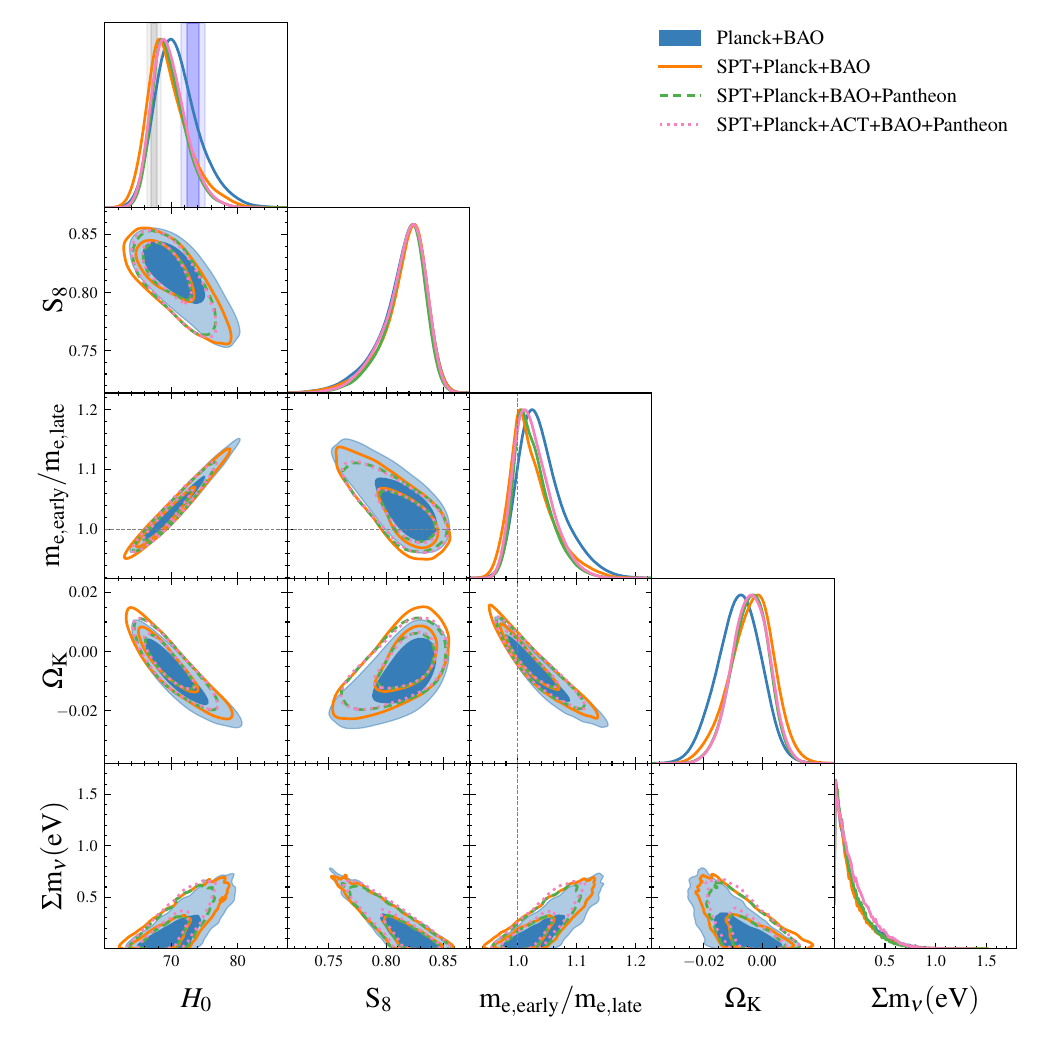}
               \caption{Same as Figure~\ref{fig:Mnu_Nsidr_Main} for the varying $m_\mathrm{e} + \Omega_k + \Sigma m_{\nu}$ model.}
               \label{fig:Me_Mnu_Omk_main}
           \end{figure}

As mentioned already, the key idea behind this model is that a variation in the electron mass allows a shift in the time of recombination and opens a new degeneracy in the space of cosmological parameters~\cite{Planck_VarMe,VarMe_Omk}. Indeed, if the last scattering surface is shifted, for instance further away from us, we need to re-adjust essentially all the cosmological parameters to keep the CMB and Large Scale Structure (LSS) spectra in agreement with observations. These re-adjustments include an increase in $H_0$ (in order to lower the angular diameter distance to recombination and compensate for the shift in the recombination time). This degeneracy has similarities with the perfect scaling transformation discussed in \cite{Cyr-Racine:2021oal} but is not exactly equivalent to it, thus still allowing us to derive bounds on the parameters of this model from CMB and LSS data. 

However, using only CMB information, the degeneracy extends extremely far in parameter space, as shown in Figure~\ref{fig:Me_SPTvsPlanck_Main}. With its limited information on large and intermediate angular scales, SPT alone is unable to accurately constrain the distance to the last scattering surface and is compatible with wide confidence intervals for $m_{\text{e,early}}/m_{\text{e,late}}$ and all other correlated parameters such as $H_0$. The posterior peaks in $m_{\text{e,early}}/m_{\text{e,late}}\simeq1.2$, corresponding to a value of $H_0$ even larger than the one measured by SH0ES. Planck alone can constrain the distance to last scattering much better, but along the direction of degeneracy, the posterior remains wide enough to encompass both the $\Lambda$CDM case ($m_{\text{e,early}}/m_{\text{e,late}}=1$, $H_0\sim67\,$km/s/Mpc) and tension-solving models ($m_{\text{e,early}}/m_{\text{e,late}}\sim 1.05$, $H_0 \sim 73\,$km/s/Mpc). One actually needs to add BAO data in order to really resolve the degeneracy and get definite predictions, for instance  $m_{\text{e,early}}/m_{\text{e,late}}=1.0046\pm0.0089$ and $H_0=67.8\pm1.1$km/s/Mpc for SPT+Planck+BAO (68\%CL), which brings us back to a tension of $Q_\mathrm{MPCL}=3.7\sigma$. In summary, most of the constraining power for this model comes from BAO data, such that the new CMB information contained in SPT and ACT data does not play a significant role. 

In ref.~\cite{H0_Olympics}, this model passed (shortly) the 3$\sigma$ threshold for all tension metrics and won a ``golden medal''. In our analysis, the prediction of this model for $H_0$ from CMB alone has hardly changed, but due to updated BAO and SH0ES measurements, the tension has grown significantly, with $Q_\mathrm{MPCL}\sim 3.7\sigma$ for any data combination that includes at least Planck and BAO data (see Figure~\ref{fig:Graph_Table3}). For the ${\cal D}_{\rm SPBP}$ data set, the model has a nearly Gaussian $H_0$ posterior and reaches $Q_{\rm DMAP}=3.9\sigma$ and $Q_\mathrm{MPCL}=3.8\sigma$, well above the $3\sigma$ threshold adopted in \cite{H0_Olympics} and in this work for considering a test as passed. Finally, the minimum $\Delta\chi^2_\mathrm{w/}$ of this model compared to $\Lambda$CDM is -18.0, which gives $\Delta$AIC$_\mathrm{w/}=-16.0$.  As previously stated, this outcome is mainly due to the improved SH0ES measurements, which make $\Lambda$CDM even less favored compared to the varying $m_e$ model. When removing the SH0ES likelihood, the model is performing as well as $\Lambda$CDM ($\Delta\chi^2_\mathrm{w/o}=0.0$), albeit with the penalty of an additional parameter ($\Delta$AIC$_\mathrm{w/o}$=2.0).

Next, we include the neutrino mass in the list of varying parameters, which results in a combination that, to our knowledge, has not been probed before. 
Interestingly, Figure~\ref{fig:Me_Mnu_Main} shows that in this case, the parameters $H_0$, $\Sigma m_\nu$ and $m_{\text{e,early}}/m_{\text{e,late}}$ are all positively correlated -- while in simpler models the correlation between $\Sigma m_\nu$ and $H_0$ is usually negative. In the absence of Planck data, that is, with SPT alone or SPT+BAO data, this model is perfectly compatible with the SH0ES measurement. However, as soon as Planck data is included, the $\Lambda$CDM+$m_e$+$\Sigma m_\nu$ model hardly improves over the $\Lambda$CDM+$m_e$ one, with tension levels reduced from $\sim3.8\sigma$ to $\sim3.5\sigma$ only. The two $\Delta$AIC metrics show that this model does not fit CMB, BAO, and SN Ia data better than $\Lambda$CDM, while in the presence of SH0ES data, it is preferred over $\Lambda$CDM at the same level as the previous model with fixed neutrino mass. Therefore, in this case, allowing the neutrino mass to vary makes no significant difference.

On the other hand, there is a well-known degeneracy between $H_0$, 
$m_{\text{e,early}}/m_{\text{e,late}}$ and $\Omega_K$ allowing to reach larger values of $m_{\text{e,early}}/m_{\text{e,late}}$ -- and thus also of $H_0$ \cite{VarMe_Omk}. The $\Lambda$CDM model features a positive correlation between $H_0$ and $\Omega_K$, due to their opposite effect on the angular diameter distance to recombination. However, with a fixed recombination time, the ``geometrical degeneracy'' between $H_0$ and $\Omega_K$ is broken by the measurement of CMB and BAO peaks at different redshifts. This is no longer the case when the recombination time also depends on a free parameter like $m_{\text{e,early}}/m_{\text{e,late}}$. In this case, the correlation between $H_0$ and $\Omega_\mathrm{K}$ even changes sign. Interestingly, compared to the results of ref.~\cite{H0_Olympics}, we find that recent SPT-3G data play a significant role in disfavoring models with the highest values of $m_{\text{e,early}}/m_{\text{e,late}}$ and $H_0$, and thus with the smallest values of $\Omega_K$. 
This is likely due to a lifting of the ``geometrical degeneracy'' by additional information contained in SPT-3G polarization data on intermediate scales.
Furthermore,  this is clearly visible in Figure~\ref{fig:Me_Omegak_main} when comparing the Planck+BAO contours with the SPT+Planck+BAO ones. With Planck+BAO data alone, this model predicts $H_0=69.1\pm 2.1\,$km/s/Mpc. Compared to $\Lambda$CDM, it reduces the Hubble tension thanks to a shift in both the mean and error on $H_0$, such that $Q_\mathrm{MPCL}=1.9\sigma$. The inclusion of SPT data reduces both the mean and error, with $H_0=67.7^{+1.9}_{-1.8}\,$km/s/Mpc, and the tension raises to $Q_\mathrm{MPCL}=2.8\sigma$. Among the models considered in this work, this is the one for which the addition of SPT data to Planck+BAO has the biggest impact. Further adding Pantheon or ACT data marginally affects the tension level.

Furthermore, ref.~\cite{H0_Olympics} found that the $\Lambda$CDM+$m_e$+$\Omega_K$ model reduces the Hubble tension roughly to the $2\sigma$ level (with $\Delta_\mathrm{GT}=2.0$ and Q$_\mathrm{DMAP}=1.9$). With the addition of SPT, eBOSS DR16, and new SH0ES data, we find that the tension is only reduced to the $\sim3\sigma$ level ($Q_\mathrm{MPCL}=2.9\sigma$ and $Q_\mathrm{DMAP}=3.1\sigma$ for $\mathcal{D}_\mathrm{SPBP}$). Still, the model has a very good metric $\Delta$AIC$_\mathrm{w/}=-20.7$. So, if we consider a $3\sigma$ tension as marginally acceptable, this model does pass our tests. In reality, this just means that $\Lambda$CDM+$m_e$+$\Omega_K$ remains marginally compatible with all our data sets. 
The model only owes its success to its very large error bar on $H_0=68.2\pm1.6$ km/s/Mpc for ${\cal D}_\mathrm{SPBP}$ and does not provide a better fit to the data in the absence of SH0ES ($\Delta$AIC$_\mathrm{w/o}=+3.0$). Thus, with current data, this model can hardly be considered an appealing explanation for the tension.

Finally, varying additionally the neutrino mass does make a difference (see Figure~\ref{fig:Me_Mnu_Omk_main}). It appears that a combined variation of the three additional parameters $m_{\text{e,early}}/m_{\text{e,late}}$, $\Omega_K$ and $\Sigma m_\nu$ together with $\Lambda$CDM parameters allows to remain in agreement with all data sets up to very large Hubble parameter values. The unusual positive correlation between $\Sigma m_\nu$ and $H_0$ is even stronger in this case than in the flat $\Lambda$CDM+$m_e$+$\Sigma m_\nu$ model. For the $\mathcal{D}_\mathrm{SPBP}$ set, the mean value of $H_0$ is significantly increased compared to the $\Lambda$CDM prediction, with $H_0 = 69.8^{+1.9}_{-2.9}$km/s/Mpc. The MPCL and DMAP metrics are relatively different from each other, consistently with the fact that the $H_0$ posterior is strongly non-Gaussian. However, they are both below the $3\sigma$ threshold, with  Q$_\mathrm{MPCL}=1.5$ and Q$_\mathrm{DMAP}=3.0$ (technically, 2.96). Thus, this model is even more compatible with all data sets when $\Sigma m_\nu$ is varied rather than fixed to $0.06\,$eV. However, in none of these two cases does the model provide a better fit to CMB, BAO, and Pantheon data, since in both cases the minimum $\chi^2$ only decreases by one unit (see $\Delta \chi^2_\mathrm{w/o}$ column in Table~\ref{Table:Summary}). When the neutrino mass is varied, the best-fit $\chi^2$ to the full data set including SH0ES is slightly lower (by $-1$) but the $\Delta$AIC$_\mathrm{w/}$ metric is slightly worse (by $+1$) due to the additional free parameter (see $\Delta$AIC$_\mathrm{w/}$ column in Table~\ref{Table:Summary} for $m_e+\Omega_\mathrm{K}$ and $m_e+\Omega_\mathrm{K}+\Sigma m_{\nu}$). Moreover, although the best-fitting value of $\Sigma m_\nu$ is close to zero for the $\mathcal{D}_\mathrm{SPBP}$ set, large values of $H_0$ close to $\sim 73\,$km/s/Mpc require relatively heavy neutrinos with $\Sigma m_\nu \sim 0.1$ to $0.3\,$eV, a slightly positive spatial curvature $K$ such that $\Omega_K\sim -0.015$ to $-0.010$, and a decrease of the electron mass between recombination and today by a few percent with $m_{\text{e,early}}/m_{\text{e,late}}\sim 1.05$ to $1.08$.

Given the precision of current data, there are not so many cosmological scenarios compatible with large values of the summed neutrino mass. This model provides a noticeable exception, with a 95\%CL upper bound $\Sigma m_\nu <0.48\,$eV for the $\mathcal{D}_\mathrm{SPBP}$ data set. This bound is three times looser than in the $\Lambda$CDM+$\Sigma m_\nu$ model, but still too strong to allow for a significant detection of the effective electron neutrino mass with the KATRIN\footnote{\url{katrin.kit.edu}} experiment~\cite{KATRIN:2005fny}.

We conclude that the $\Lambda$CDM+$m_e$+$\Omega_K$+$\Sigma m_\nu$ model passes the MPCL, DMAP, and AIC (with SH0ES) tests. While it is still not a convincing explanation of the tension, since it does not improve the goodness-of-fit to the data in the absence of SH0ES, it would still be interesting to investigate with future data, e.g. with improved CMB polarization measurements.




 \subsection{Early Dark Energy}
 \label{SubSec:EDE_Results}

 \begin{figure}[hbt!]
                \centering
                \includegraphics[width=15cm,angle=0]{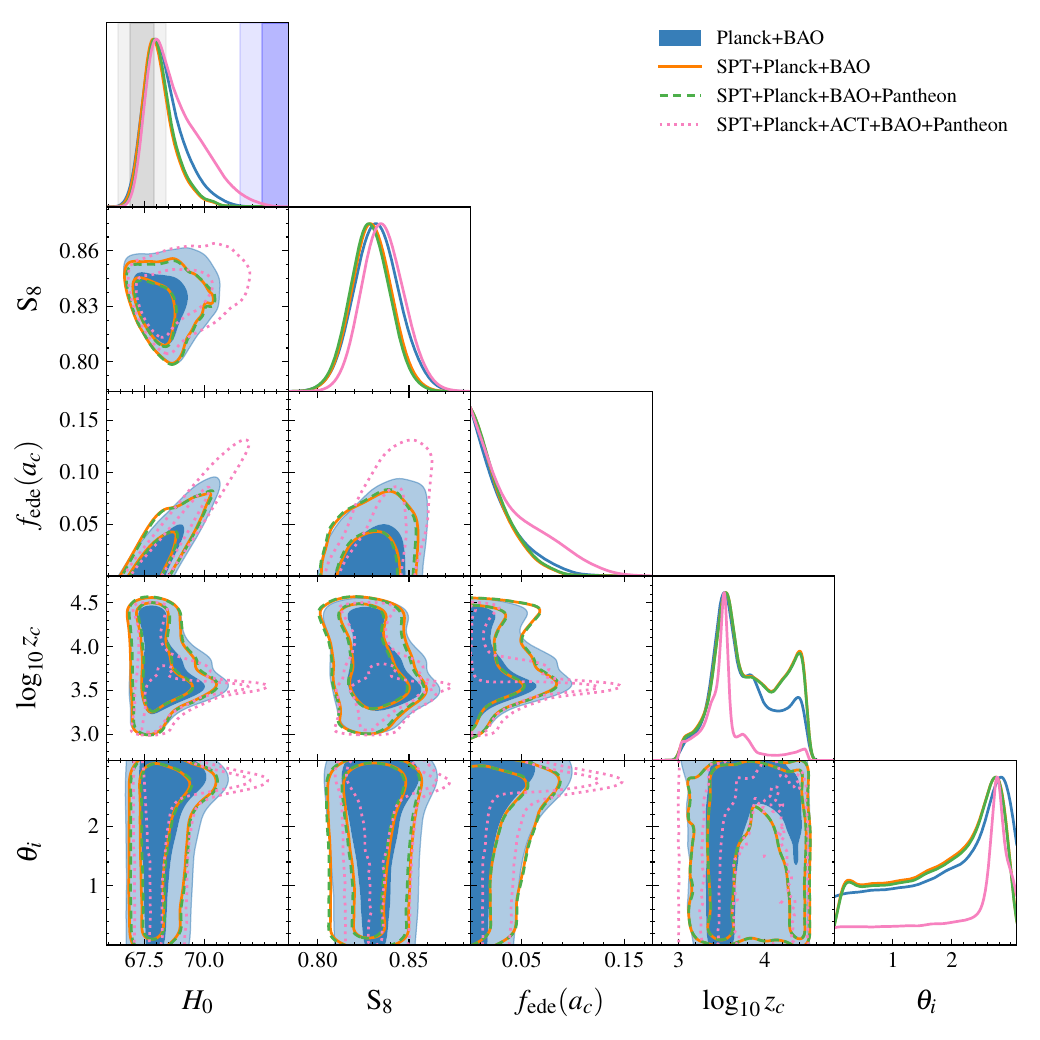}
                \caption{Same as Figure \ref{fig:Mnu_Nsidr_Main} for the EDE model.}
                \label{fig:EDE_Main}
            \end{figure}
            
\begin{figure}[hbt!]
                \centering
                \includegraphics[width=15cm,angle=0]{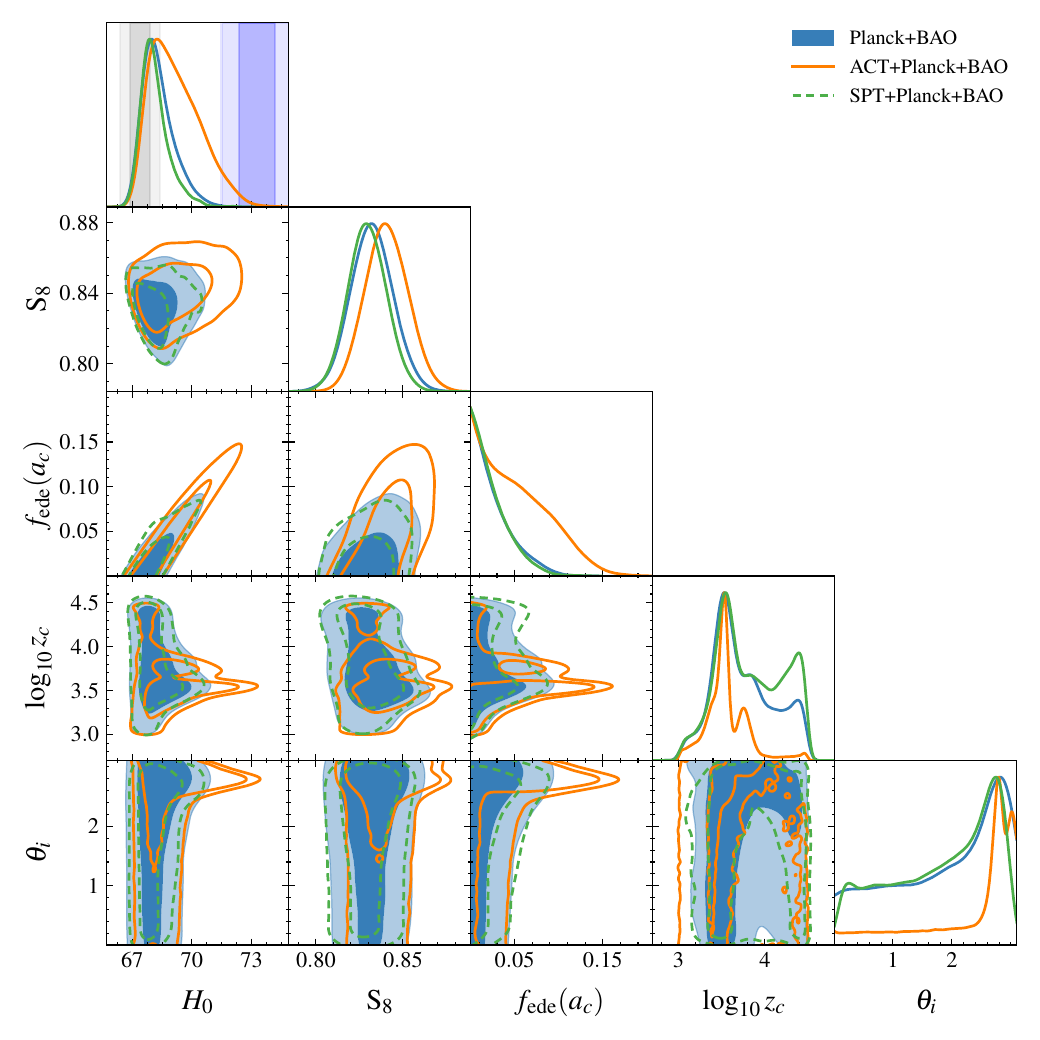}
                \caption{Same as Figure \ref{fig:EDE_Main} with a focus on the impact of SPT data compared to that of ACT.}
                \label{fig:EDE_SPTvsACT_Main}
            \end{figure}
            
A few interesting features appear for this model when looking at Table~\ref{Table:H0_values}. First, the strongest constraints come mainly from CMB data, particularly from Planck. Comparing $\mathcal{D}_\mathrm{SP}$ and $\mathcal{D}_\mathrm{SPB}$, we see only a slight decrease in error bars, with the mean value barely changing. 

Second, we see that the combination ${\cal D}_\mathrm{SPB}$ puts significantly tighter constrains on the Hubble rate ($H_0=68.14^{+0.43}_{ -0.78}$ km/s/Mpc) than ${\cal D}_\mathrm{PB}$ ($H_0=68.3^{+0.52}_{ -0.98}$ km/s/Mpc). This sounds paradoxical at first sight, because SPT data alone or even SPT+BAO data are not very constraining on $H_0$ compared to Planck+BAO, and even suggest a higher mean value than Planck+BAO. However, the very same behavior was noticed and documented in \cite{Vivian_Tristan_EDE} (fourth-to-last paragraph). One can get more insight into this by comparing the 2D confidence contours on all parameters in the Planck+BAO and SPT+BAO cases. The later contours are overall wider, and usually encompassing the former. However, the SPT+BAO data set is better fitted by higher values of $\log_{10}z_c$ and smaller values of $S_8$ (or equivalently $A_s$) than the Planck+BAO data. In this particular parameter space, the contours only have a small overlap in a region which turns out to be consistent only with small values of $H_0$. This explains why the combined data set is more constraining than each individual data set, even if the individual ones are statistically compatible with each other.

Overall, the combination of recent data reduces the ability of the EDE model to reach large values of $H_0$ compared to older analyses. For our reference ${\cal D}_\mathrm{SPBP}$ data set, the MPCL metric gives $Q_\mathrm{MPCL}=3.7\sigma$. However, the DMAP metric conveys a slightly different message, since $Q_\mathrm{DMAP}=2.7\sigma$. The fact that these two numbers differ significantly is consistent, first, with the strong non-Gaussianity of the $H_0$ posterior for this model, and second, with the well-known ``prior volume effect'' that frequently occurs in Bayesian parameter estimation~\cite{Prior_Volume1,Prior_Volume2}.

This effect is particularly important for models in which one parameter accounts for the presence (or absence) of an ingredient, which is not required by the data with high significance. In our case, this parameter is the EDE fraction, $f_\mathrm{EDE}$, whose posterior is compatible with zero.  When $f_\mathrm{ede}\rightarrow 0$, the EDE field has no observable impact, i.e. we are back to $\Lambda$CDM, but the parameters $\theta_\mathrm{i}$ and $z_c$ can still take any possible value. We see in Figure~\ref{fig:EDE_Main} that many models within the confidence contours have a very small $f_\mathrm{EDE}$ and a wide range of possible values for the EDE parameters $\theta_i$ and $\log_{10}z_c$. Large values of $H_0$ reducing the Hubble tension require specific values of $(\theta_i,\log_{10}z_c)$, close to (2.8, 3.5), but in the $H_0$ posterior, smaller values of $H_0$ are strongly enhanced by the marginalisation over other values of $(\theta_i,\log_{10}z_c)$, which are unconstrained in the $f_\mathrm{EDE} \rightarrow 0$ (or equivalently $\Lambda$CDM) limit. In order to avoid this effect, several works switched to grid sampling or profile likelihood studies~\cite{Profile_Likelihood_EDE1,Elisa-Eiishiro,Elisa-Laura} (see~\cite{Ups_Downs_EDE} for more detailed discussion).
 
In summary, the EDE model has a reasonably small $Q_\mathrm{DMAP}$ value because the best fit to the full data set returns a relatively good $\chi^2$. On the other hand, we find a much larger $Q_\mathrm{MPCL}$ because the Bayesian $H_0$ posterior is enhanced for small values of $H_0$ by the marginalisation over allowed models close to $\Lambda$CDM. In this case, giving more credit to one or the other metric becomes a matter of taste. The Bayesians will stress the fact that EDE models providing a good fit to all data sets are unlikely. The frequentists will instead give credit to the observation that there exist some parameter values for the EDE model such that the tension gets reduced to the $2.7\sigma$ level. Once more, we see that different metrics play a complementary role. Here, we remain agnostic and we consider that models for which one of the MPCL or DMAP metrics remains below the $3\sigma$ level provide an acceptable reduction of the tension.

It should also be noted that, since $Q_\mathrm{MPCL}$ is strongly affected by Bayesian volume effects, we would find different values of this estimator if we adopted different top-hat prior edges on the parameters $(\theta_i,\log_{10}z_c)$: if we had some particular reasons to assume narrower priors, $Q_\mathrm{MPCL}$ would shrink and possibly become as small as $Q_\mathrm{DMAP}$.

Looking at the goodness-of-fit of the EDE model compared to $\Lambda$CDM, we find that the EDE model is the only one in our list reducing the best-fit $\chi^2$ to all-but-SH0ES data by more than one unit, with $\Delta \chi^2_\mathrm{w/o}=-4.6$. However, this provides no hint that the model is a significantly better fit to all-but-SH0ES data because the EDE model has three additional free parameters and a positive $\Delta$AIC$_\mathrm{w/o}=+1.4$. However, this ability to reduce slightly the best-fit $\chi^2$ to CMB, BAO, and supernovae data leads to the best AIC among all our considered models when SH0ES data is included, with $\Delta$AIC$_\mathrm{w/}=-25.1$. Once more, one should take this result with a grain of salt, since it mainly reflects the high level of tension of the $\Lambda$CDM case.

We finally consider the inclusion of ACT-DR4 data~\cite{ACT1,ACT2,ACT_EDE}. Like several previous works (see~\cite{Ups_Downs_EDE} and citations within), we find that these data are slightly more compatible with large values of the EDE fraction. In Figure~\ref{fig:EDE_SPTvsACT_Main}, we show the result of an additional run with Planck+ACT+BAO data, in which a value of $f_\mathrm{ede}\approx0.1$ is excluded at the $<2\sigma$ level for Planck+ACT+BAO, while it is at more than $3\sigma$ when ACT is not added. In the same Figure, one can look at the contours in the $(\theta_i,\log_{10}z_c)$ space. With Planck and SPT data, these parameters are loosely constrained, since most of the allowed models lay very close to the $\Lambda$CDM limit. Thus, the contours have an extended and complicated shape. In the presence of ACT data, the contours narrow down around $(\theta_i,\log_{10}z_c)\sim (2.8,3.5)$, that is, around the values for which a fraction of EDE can contribute to the expansion rate before photon recombination, becoming more compatible with high values of $H_0$. When Planck, SPT, and ACT are taken into account together, SPT and ACT data have marginally antagonist effects on the $f_\mathrm{EDE}$ posterior and the fit favors a compromise region in parameter space. The MPCL metric gets reduced from 3.7 to 3.1 when adding ACT data to ${\cal D}_\mathrm{SPB}$. Thus, like previous authors, we find that SPT-3G and ACT-DR4 data contain some slightly contradictory information on high-$\ell$ multipoles that push the allowed fraction of EDE in opposite directions, such that this model would appear as a better solution to the Hubble tension if ACT was trusted more than SPT, and vice versa. It will be interesting to update this comparison when new ACT data will be released.

 \subsection{Majoron}


\begin{figure}[hbt!]
        \centering
        \includegraphics[width=15cm,angle=0]{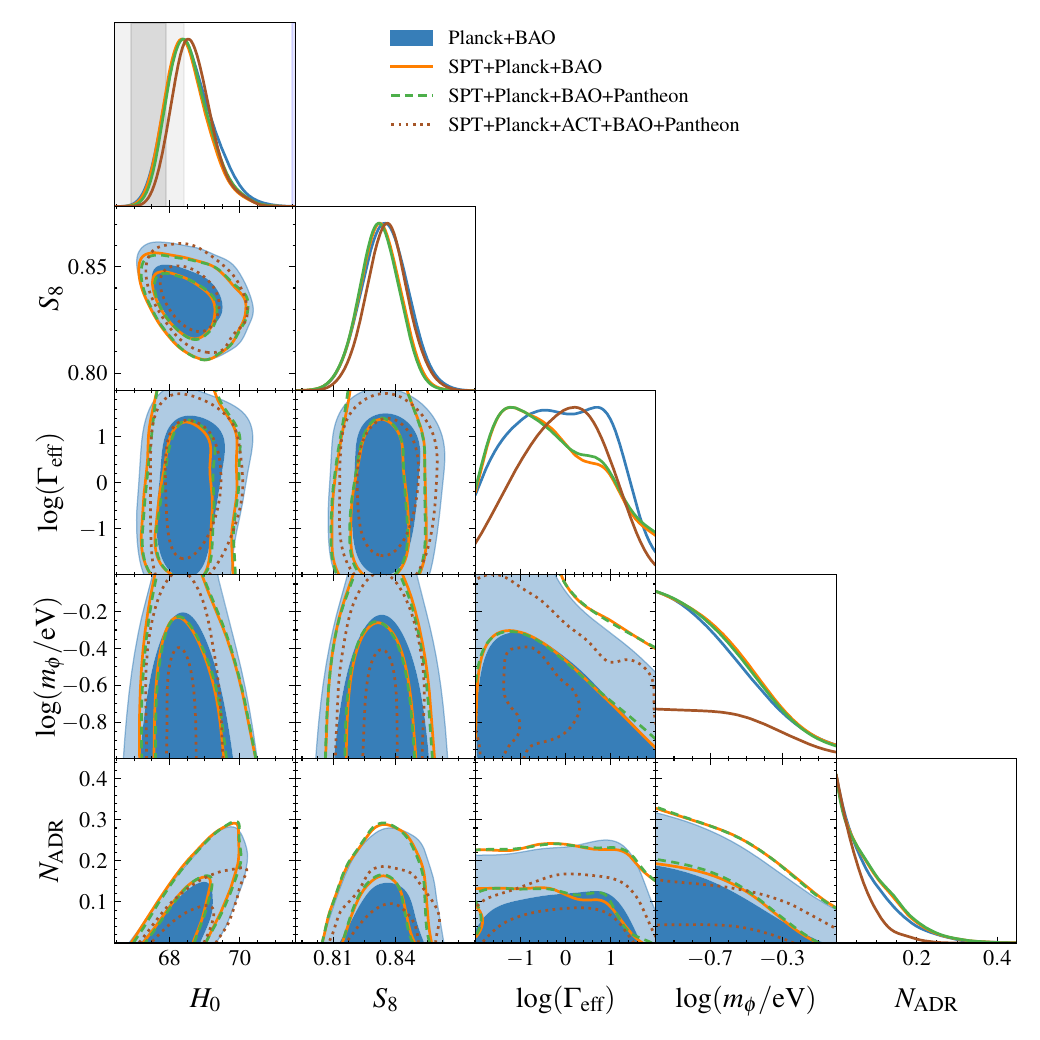}
        \caption{Same as Figure \ref{fig:Mnu_Nsidr_Main} for the Majoron model, with a prior $m_\phi<1\,$eV to single out the small-mass scenario.}
        \label{fig:maj_sub_main}
\end{figure}

\begin{figure}[hbt!]
        \centering
        \includegraphics[width=15cm,angle=0]{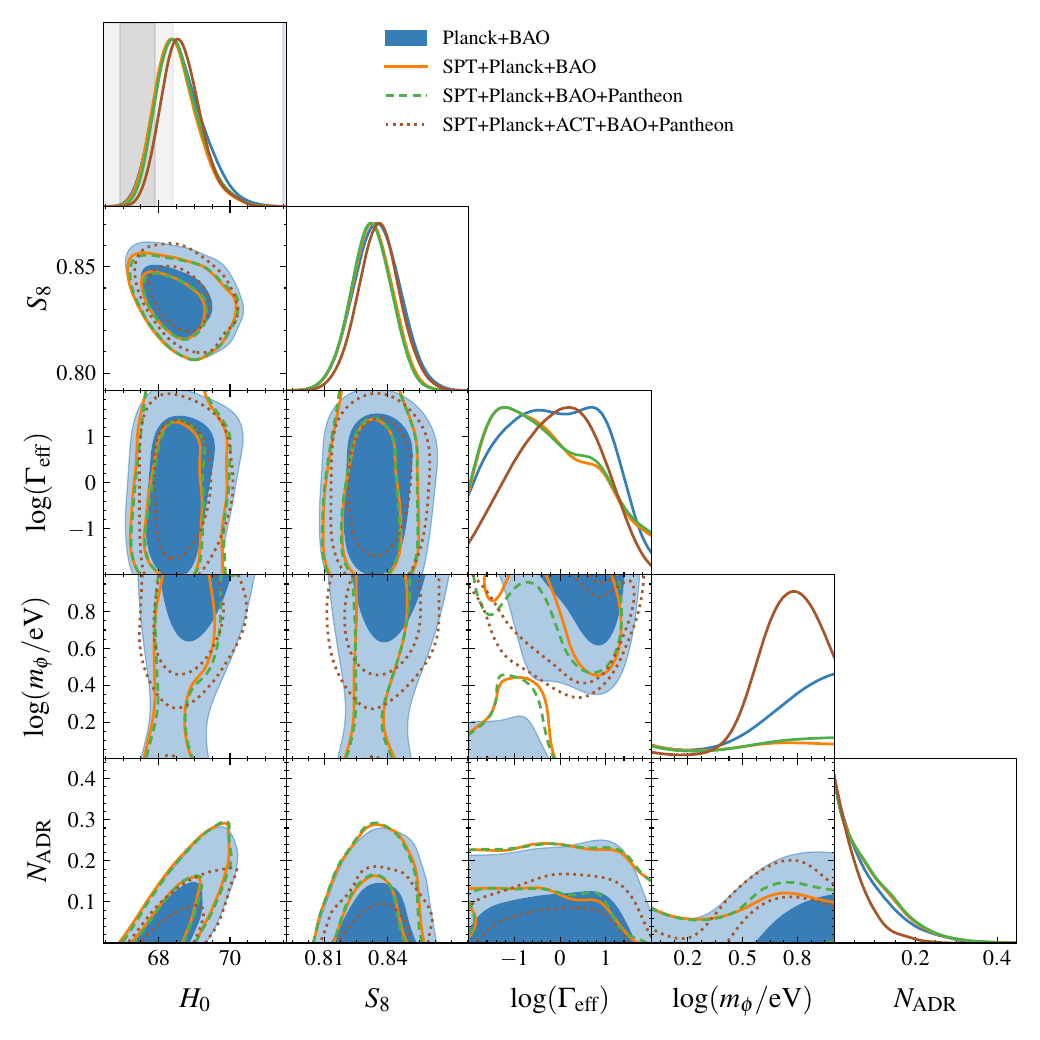}
        \caption{Same as Figure \ref{fig:maj_sub_main} with a prior $m_\phi>1\,$eV to single out the large-mass scenario.}
        \label{fig:maj_sup_main}
\end{figure}

The first thing to be noticed about the Majoron model is that, when fitted to data combinations including at least Planck and BAO, it may return a bi-modal posterior, with one preferred region corresponding to a Majoron mass $m_{\phi}<1\,$eV and another one to $m_{\phi}>3\,$eV. Bi-modal posteriors are notoriously difficult to sample with the Metropolis-Hastings algorithm. We circumvent this problem by presenting separately the posteriors and contours obtained
with a prior $-1 \leq \log(m_\phi /{\rm eV}) \leq 0$ (see Figure~\ref{fig:maj_sub_main}) or $0 \leq \log(m_\phi /{\rm eV}) \leq 1$ (see Figure~\ref{fig:maj_sup_main}). In each case, the posterior is unimodal.

To assess the relevance of each region of parameter space, we can simply compare the minimum $\chi^2$ within the MCMC for each of the two priors. (The  minimum $\chi^2$'s from MCMC chains are not as stable and robust as the ones we obtained for previous models using a specific minimisation algorithm, but they are indicative enough for the purpose of this section.) For Planck and BAO alone, the two cases have a very similar minimum $\chi^2$ and the corresponding regions in parameter space can be roughly considered as equally likely. However, when adding SPT and then SNIa data, we find that the low-mass region is preferred. Thus, in the various tables, we report the results based on this case. Note that the Hubble parameter does not correlate with $m_\phi$, such that the $H_0$ posterior is very similar in the two cases.

To gain some physical insight into this model, we can think of it as effectively equivalent to a combination of the $\Lambda$CDM+$N_{\rm eff}$ and $\Lambda$CDM+$N_{\rm SIDR}$ ones, with additional parameters tuning the evolution of the density of both free-streaming and self-interacting relativistic relics as a function of time.
Thus, it is natural to compare the performance of this model to the former ones. We should however keep in mind that in previous sections, we varied $N_{\rm eff}$ or $N_{\rm SIDR}$ in combination with $\Sigma m_\nu$, while here we neglect neutrino masses for simplicity.\footnote{As mentioned in section~\ref{Sec:Models}, we use a version of {\tt CLASS} developed by~\cite{Sandner:2023ptm} that does not include yet the case in which the Majoron interacts with massive neutrinos.} However, in the $\Lambda$CDM+$\Sigma m_\nu$+$N_{\rm eff}$ and $\Lambda$CDM+$\Sigma m_\nu$+$N_{\rm SIDR}$ models, we did not observe a correlation between the radiation density parameter and the summed neutrino mass. Thus, it is still fair to compare the Majoron scenario to the former models. However, we should not compare the posterior of $N_{\rm ADR}$ directly with that of $\Delta N_{\rm eff}$ or $N_{\rm SIDR}$, because even in the absence of additional dark radiation ($N_{\rm ADR}=0$) the Majoron scenario already predicts an enhancement of $N_{\rm eff}$ with respect to the standard value 3.044, which is caused, first, by the density of Majoron produced by inverse neutrino decay, and later on, by that of additional active neutrinos produced by the Majoron decay.

One can check in Figure~\ref{fig:Graph_Table3} that depending on the data set, the Majoron model performs either a bit better or a bit worse than the $\Lambda$CDM+$\Sigma m_\nu$+$N_{\rm eff}$ and $\Lambda$CDM+$\Sigma m_\nu$+$N_{\rm SIDR}$ models as far as the reduction of the tension is concerned. However, it does not bring any decisive improvement.
%
%
With the ${\cal D}_\mathrm{PB}$ data set, we find $Q_\mathrm{MPCL}=4.0\sigma$, which is significantly worse than the result obtained in~\cite{H0_Olympics} also with Planck+BAO data, $\Delta_\mathrm{GT}=2.7\sigma$. This degradation is not only caused by the more recent BAO and SH0ES data used in the current analysis but also by the more accurate modeling of the Majoron scenario (based on the work of refs.~\cite{Nu_Daming1,Nu_Damping2}), in the updated branch of {\tt CLASS} developed by the authors of  \cite{Sandner:2023ptm} and used in this work. Additionally, we find that the tension slightly increases when adding SPT data, with $Q_\mathrm{MPCL}=4.2\sigma$ for ${\cal D}_\mathrm{SPB}$. For our reference data set, ${\cal D}_\mathrm{SPBP}$, we find $Q_\mathrm{MPCL}=4.3\sigma$. One may notice that, like in the EDE case, the situation improves a bit when including ACT-DR4 data, with $Q_\mathrm{MPCL}=4.0\sigma$ for ${\cal D}_\mathrm{SPAB}$. However, when adding also Pantheon data, we are back to $Q_\mathrm{MPCL}=4.4\sigma$. We conclude that the Majoron model is no longer able to bring the Hubble tension down to acceptable values, and we do not compute other tension metrics for this model.

\subsection{Update on the $S_8$ Tension}
\label{sec:S_8_Tension}

     \begin{table}[hbt!]
         \resizebox{1.1\textwidth}{!}{%
             \hspace{-2.5cm}
             \renewcommand{\arraystretch}{1.5}
             \begin{tabular} {|l |c c c c c >{\bf}c c c|}
             \hline
             Models &  $\mathcal{D}_{\text{S}}$ & $\mathcal{D}_{\text{SP}}$ & $\mathcal{D}_{\text{SB}}$ & $\mathcal{D}_{\text{PB}}$ & $\mathcal{D}_{\text{SPB}}$ & $\mathcal{D}_{\text{SPBP}}$& $\mathcal{D}_{\text{SPAB}}$ & $\mathcal{D}_{\text{SPABP}}$\\
             \hline
             {$\Lambda$CDM} & 0.798 $\pm$ 0.042 & 0.830 $\pm$ 0.012 & 0.819 $\pm$ 0.017 & 0.830 $\pm$ 0.012 & 0.828 $\pm$ 0.010 & 0.826 $\pm$ 0.010 & 0.832 $\pm$ 0.009 & 0.830 $\pm$ 0.010\\

             \hline
             {+$\Sigma m_{\nu}$} & 0.772 $\pm$ 0.045 & $0.830\pm 0.012$ & 0.776 $\pm$ 0.037 & 0.828 $\pm$ 0.011 & 0.827 $\pm$ 0.011 & 0.826 $\pm$ 0.011 & 0.829 $\pm$ 0.010 & 0.829 $\pm$ 0.010\\

             \hline
             {+$\Sigma m_{\nu}$+CPL} & 
             0.761 $\pm$ 0.060 & 0.774 $\pm$ 0.033 & 0.779 $\pm$ 0.039 & 0.837 $\pm$ 0.014 & 0.834 $\pm$ 0.014 & 0.826 $\pm$ 0.012& 0.835 $\pm$ 0.016 & 0.827 $\pm$ 0.012\\
             \hline
             {+$\Sigma m_{\nu} +$ N${}_{\text{eff}}$} & 0.772 $\pm$ 0.048 & $0.830\pm 0.012$ & 0.785 $\pm$ 0.044 & 0.831 $\pm$ 0.011 & 0.828 $\pm$ 0.012 & 0.828 $\pm$ 0.011 & 0.830 $\pm$ 0.010 & 0.829 $\pm$ 0.010\\
             \hline
             {+$\Sigma m_{\nu}$+ N$_{\text{SIDR}}$} & 0.769 $\pm$ 0.049 & 0.825 $\pm$ 0.012 & 0.780 $\pm$ 0.042 & 0.827 $\pm$ 0.011 & 0.825 $\pm$ 0.011 & 0.825 $\pm$ 0.011 & 0.828 $\pm$ 0.010 & 0.828 $\pm$ 0.010\\

             \hline
             {+$\Sigma m_{\nu}+ \Omega_{\text{k}}$} & --- & 
             0.873 $\pm$ 0.025 & 0.763 $\pm$ 0.041 & 0.829 $\pm$ 0.011 & 0.827 $\pm$ 0.011 & 0.826 $\pm$ 0.011 & 0.824 $\pm$ 0.011 & 0.828 $\pm$ 0.010 \\
             \hline
             {$m_e$}  & 0.687 $\pm$ 0.155& $0.943\ \pm\ 0.081$ & $0.806\ \pm \ 0.029$ & $0.829\ \pm \ 0.010$ & $0.828\ \pm \ 0.010$ & 0.827 $\pm$ 0.010 & $0.830\ \pm \ 0.010$ & $0.829\ \pm \ 0.009$\\
             \hline
             {$m_e$+$\Sigma m_{\nu}$} & 0.727 $\pm$ 0.071 & 0.873 $\pm$ 0.040 & 0.709 $\pm$ 0.064 & 0.827 $\pm$ 0.012 & 0.824 $\pm$ 0.013 & 0.824 $\pm$ 0.012 & 0.825 $\pm$ 0.013 & 0.824 $\pm$ 0.013\\

             \hline
             {$m_e$ + $\Omega_k$} & 0.770 $\pm$ 0.062 & $0.878 \pm 0.041$ & $0.810 \pm 0.030$ & $0.828 \pm 0.010$ & $0.828 \pm 0.010$ & 0.827 $\pm$ 0.010 & $0.830 \pm 0.010$ & $0.828 \pm 0.010$\\
             \hline
             {$m_e$ + $\Omega_k$ + $\Sigma m_{\nu}$} & --- & --- 
             &0.731 $\pm$ 0.072 & 0.814 $\pm$ 0.021 & 0.816 $\pm$ 0.020 & 0.816 $\pm$ 0.018 & 0.816 $\pm$ 0.021 & 0.816 $\pm$ 0.019 \\
             \hline
             {EDE} & 
             0.797 $\pm$ 0.044 & 0.832 $\pm$ 0.013 & 0.819 $\pm$ 0.023 & 0.832 $\pm$ 0.012 & 0.830 $\pm$ 0.011 & 0.829 $\pm$ 0.011 & 0.837 $\pm$ 0.012  & 0.836 $\pm$ 0.012 \\
             \hline
             {Majoron} & 0.792 $\pm$ 0.044 & 0.829 $\pm$ 0.012 & 0.847 $\pm$ 0.022 & 0.835 $\pm$ 0.011 & 0.832 $\pm$ 0.010 & 0.832 $\pm$ 0.010 & 0.837 $\pm$ 0.010 & 0.836 $\pm$ 0.010\\

             \hline
             \end{tabular}}
             \caption{Mean values and 68\%CL credible interval of $\mathrm{S}_8$ for each model and each combination of data. The values for the reference data set $\mathcal{D}_\mathrm{SPBP}$ are highlighted in bold.}
             \label{Table:sigma8_values}
     \end{table}

    \begin{figure}[ht!]
        \centering
        \includegraphics[width=13cm,angle=0]{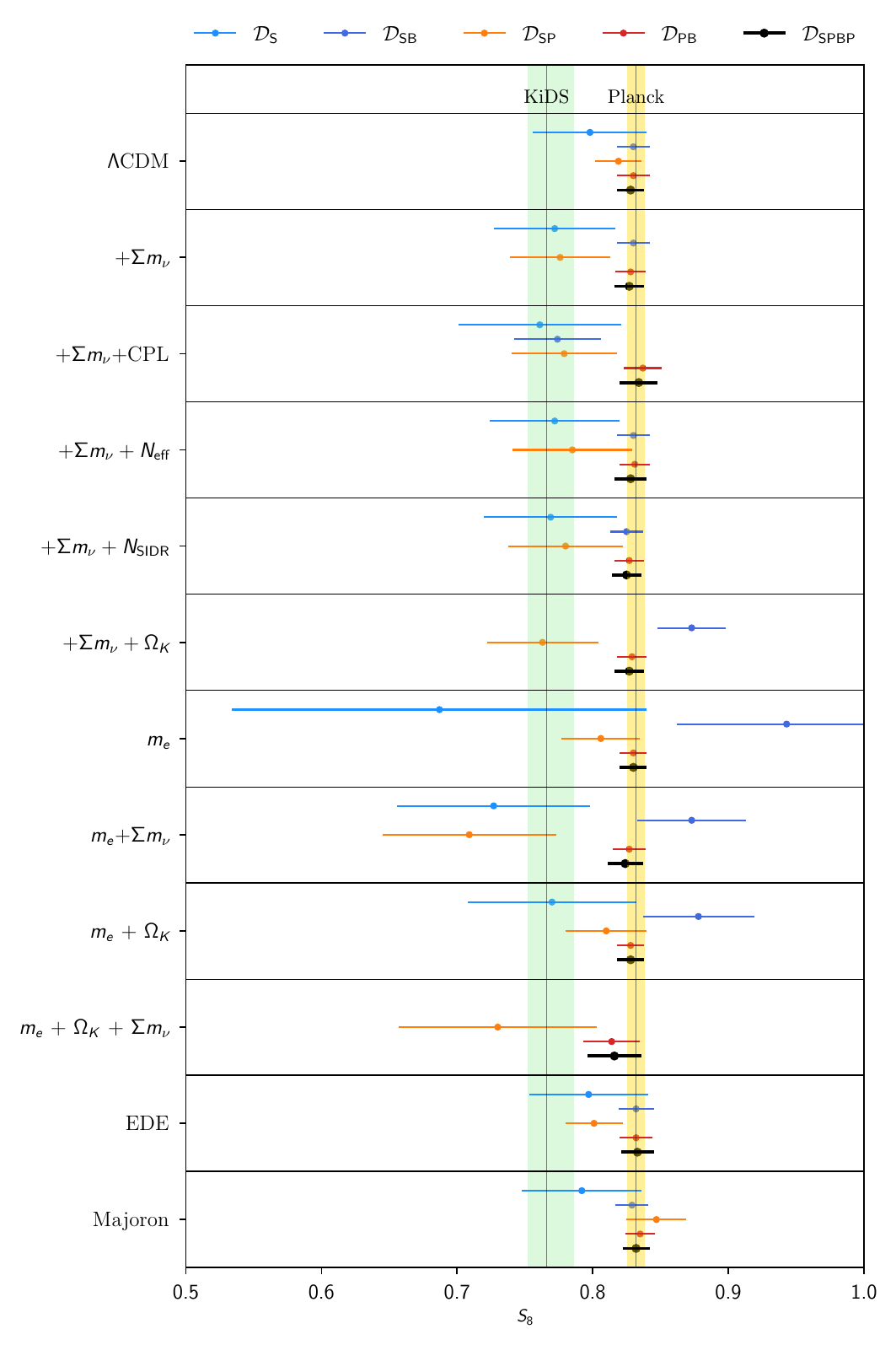}
        \caption{Graphical representation of the first four columns of Table~\ref{Table:sigma8_values}, as well as our baseline data set $\mathcal{D}_\mathrm{SPBP}$. In particular, we  highlight the difference in constraining power between Planck and SPT-3G 2018 for each studied model.}
        \label{fig:Graph_Table6}
\end{figure}

Since it is also important to check that models ease the Hubble tension without worsening the $\mathrm{S}_8$ one, we computed the Gaussian tension on $\mathrm{S}_8$ for each model and data set. The mean value of $S_8$ and its credible intervals are reported in Table~\ref{Table:sigma8_values}, with a graphical representation in Figure~\ref{fig:Graph_Table6}. 

Different weak lensing surveys and analysis pipelines derive bounds on $S_8$ that are compatible with each other but correspond to significantly different tension levels. 
For instance, one can use the KiDS measurement $S_8=0.766^{+0.020}_{-0.014}$ (68\%CL, ~\cite{KiDS}) as a reference and estimate the $S_8$ tension using the Gaussian tension metric. Then, we find that for all the models we study, including $\Lambda$CDM, the tension remains below $3\sigma$ for the  $\mathcal{D}_\mathrm{SPBP}$ data set. Indeed, in this case, $\Lambda$CDM is in $2.7\sigma$ tension, while the highest value ($2.9\sigma$) is reached by the Majoron model. Note also that the value $S_8=0.766$ reported by KiDS is a mean posterior value obtained from an asymmetric posterior. In such a case, the Gaussian tension metric should be taken with caution. 

On the other hand, one could use instead the fiducial analysis of the joint DES-Y3 plus KiDS-1000 data, reported in the first line of Table~4 of ref.~\cite{Kilo-DegreeSurvey:2023gfr}. For our purpose of estimating the Gaussian tension, we should ideally use the maximum of the $S_8$ posterior as a reference value. The joint DES+KiDS analysis provides such statistics and reports $S_8=0.792^{+0.017}_{-0.018}$ (68\%CL, Table 4 of~\cite{Kilo-DegreeSurvey:2023gfr}, column ``Maximum Marginal''). This corresponds to a 1.7$\sigma$ Gaussian tension with respect to ${\cal D}_\mathrm{SPBP}$ for $\Lambda$CDM, while it is $2\sigma$ for the Majoron model. In this case, there is essentially no $S_8$ tension in any of the studied models. Given these differences, we refrain from stating explicitly an $S_8$ tension level for each model.


 The level of $S_8$ tension reported here should be regarded as indicative only. Indeed, the $S_8$ measurements reported by DES or KiDS were obtained under the assumption of a simple cosmology (namely, $\Lambda$CDM, including a free neutrino mass in some cases). In principle, for a more accurate statement, $S_8$ should be inferred from raw weak lensing data for each of the extended cosmologies considered here. This goes far beyond the scope of this paper.

\section{Conclusions}
\label{Sec:Conclusion}

This work presents updated constraints from cosmological data on several extensions of the minimal $\Lambda$CDM model. These extensions feature combinations of physical ingredients such as neutrino masses, spatial curvature, dynamical dark energy, extra relativistic relics with different interaction properties (free-streaming, self-interacting), interactions in the active neutrino sector as predicted by a Majoron scenario~\cite{Majoron1,Majoron2,Majoron3}, a time-varying electron mass~\cite{VarMe1,Planck_VarMe,VarMe_Omk}, or Early Dark Energy~\cite{EDE1,EDE2,EDE3}. We derive constraints on the additional parameters of these models, and we focus on various approaches to quantify their potential to solve the Hubble tension in light of recent SH0ES results~\cite{New_H0}. 
Compared to previous studies~\cite{H0_Olympics,In_Realm}, our work contains three new features. Firstly, we included more recent data sets, like CMB temperature and polarization data from SPT-3G 2018 TT/ TE/EE~\cite{Ups_Downs_EDE,Recent_EDE_SPT}, additional BAO data from eBOSS DR16~\cite{SDSS_DR16}, and updated SH0ES data~\cite{New_H0}. In particular, the scenarios considered in this work had not been confronted before to the new SPT-3G data -- except for EDE, see~\cite{Ups_Downs_EDE,Recent_EDE_SPT,Vivian_Tristan_EDE}. Secondly, we studied the effect of curvature, extra relativistic relics, dynamical dark energy, or a time-varying electron mass in the framework of a baseline $\Lambda$CDM+$\Sigma m_\nu$ scenario with arbitrary neutrino mass, while previous works fixed the total mass to its minimum allowed value ($\Sigma m_\nu=0.06\,$eV), which is arbitrary. Thirdly, we proved that we are able to obtain consistent results in a much shorter time when performing our Bayesian parameter inference with an Einstein-Boltzmann code emulator~\cite{günther2023uncertaintyaware} that trains itself on the fly for each model and data set during a standard MCMC run.


We found that classical extensions of $\Lambda$CDM featuring combinations of massive neutrinos, spatial curvature, dynamical dark energy, or extra relativistic relics are unable to ease the Hubble tension in light of new data. In particular, the self-interacting dark radiation model, which obtained a ``bronze medal'' in the previous ranking proposed in \cite{H0_Olympics}, is now unable to bring the tension below the 4$\sigma$ level. This degradation is driven by new BAO and SH0ES data and enhanced by ACT data. Given their plausible physical motivations, these models would still be interesting to study on their own, assuming that the Hubble tension vanishes or can be explained with another ingredient. However, they cannot be considered a viable solution to the Hubble tension.

We then switched to the study of more elaborate scenarios that have been recently advertised as possible solutions to the Hubble tension. In the latter category, we found that assuming a time-varying electron mass (in the framework of the minimal $\Lambda$CDM or of the $\Lambda$CDM+$\Sigma m_\nu$) decreases the tension level but is no longer able to bring it below the 3$\sigma$ level: this model loses the ``silver medal'' awarded in \cite{H0_Olympics}. It was already well-known that adding spatial curvature as an additional ingredient is particularly relevant in this case because of an interesting parameter degeneracy between $H_0$, $m_\mathrm{e,early}/m_\mathrm{e,late}$ and $\Omega_K$. We confirm this, and we also find for the first time that varying the neutrino mass is also relevant in this context since the degeneracy is significantly stronger in the presence of massive neutrino effects. Thus the models with varying $m_e+\Omega_\mathrm{K}$ and varying $m_e+\Omega_\mathrm{K}+\Sigma m_{\nu}$ reduce the Hubble tension to a statistically significant level, respectively below the 3$\sigma$ and 2$\sigma$ levels (see Figure~\ref{fig:Graph_Table3}). However we should stress that the success of these models comes mainly from increasing the error bar on $H_0$, with highly non-Gaussian posterior distributions (see figures~\ref{fig:Me_Omegak_main} -\ref{fig:maj_sub_main}), rather than shifting the mean of the posterior to larger values of the Hubble parameter close to the mean SH0ES value (although the varying $m_e+\Omega_\mathrm{K}+\Sigma m_{\nu}$ does raise the predicted mean value of $H_0$ close to $70\,$km/s/Mpc, see tables~\ref{Table:Summary} and~\ref{Table:H0_values}). Finally, the EDE model is also able to reduce the Hubble tension below the $3\sigma$ level as long as we take into account the frequentist DMAP metric (in this case, the Bayesian MPCL metric is penalised by strong prior volume effects, as already discussed in previous works~\cite{Vivian_Tristan_EDE}). However, like 
in the varying $m_e+\Omega_\mathrm{K}$ case, the EDE model owes its success to larger error bars on $H_0$ rather than to an increase in the mean predicted $H_0$ value, unless ACT-DR4 data are also included in the analysis.

We conclude that the varying $m_e+\Omega_\mathrm{K}$(+$\Sigma m_\nu$) and EDE models are still valid candidates to solve the Hubble tension. Of course, we make such a statement on a purely phenomenological and cosmological-data-driven basis. We should remind the reader that, if taken seriously, these models would raise several theoretical challenges from the point of view of high-energy physics or early-universe physics. Discussing these challenges is far beyond the scope and spirit of this paper. Besides, from a purely phenomenological perspective, we have seen that, even if these models can reduce the Hubble tension, they appear as {\it ad-hoc} solutions, since there is no hint in their favor apart from the Hubble tension itself, and {\it unpredictive} solutions, since they accommodate rather than predict a large $H_0$. Thus, we clearly need more data and perhaps further theoretical modeling to give them a final assessment.

In the future, we will be revisiting these models and other elaborate ones
with upcoming data from SPT-3G and ACT. We already saw in this work that SPT-3G 2018 data brings very interesting constraints, but there is even much more information to expect from forthcoming releases~\cite{SPT-3G:2021,SPT-3G:2022}. 
Having now validated our on-the-fly emulator, we will be able to update constraints and assess Hubble tension levels for a growing number of models with a decreased amount of computational effort.


\section*{Acknowledgments}

We would like to thank Miguel Escudero,  Stefan Sandner, Nils Sch\"{o}neberg, Tristan Smith, Vivian Poulin, and Sam Witte for interesting discussions and their modified versions of \texttt{CLASS}. We also thank Erik De la Haye, Christian Fidler, and Markus Mosbech for their useful comments. We thank as well Lennart Balkenhol, Federica Guidi, Etienne Camphius and Eric Hivon from the SPT collaboration at the Institut d'Astrophysique de Paris (IAP) for fruitful conversations. This work has made use of the Infinity Cluster hosted by IAP. Sven Günther acknowledges support from the DFG grant LE 3742/6-1. Simulations were performed with computing resources granted by RWTH Aachen University under projects rwth1302, rwth1305, and thes1348. This work has received funding from the French Centre
National d’Etudes Spatiales (CNES). This project has received funding from the
European Research Council (ERC) under the European Union’s Horizon 2020
research and innovation programme (grant agreement No 101001897)

\section*{Appendices}
\appendix

\section{Accelerating MCMC runs using an online learning emulator}
\label{APP: Pipeline_Plus}


In this appendix, we provide further details about two cases where the emulator was used for the analysis pipeline of this work. In appendix \ref{sec:emulator}, we discuss the performance of the emulator-assisted posterior inference and illustrate its accuracy for two examples of MCMC analyses. In appendix \ref{subsec:minimization}, we outline our approach to find best-fit models by performing a $\chi^2$ minimization. For the first task, we find that the emulator speeds up the pipeline by at least one order of magnitude, and for the second task, by two orders of magnitude.
        
\subsection{Comparing Emulator Results with usual MCMC}
\label{sec:emulator}

In ref.\cite{günther2023uncertaintyaware}, it was shown that the emulator can reliably overcome the bottleneck of Bayesian analysis, which is usually the execution of the Boltzmann code. We confirm this in the present context. We find that, when using the emulator, the total run time is dominated by the evaluation of the CMB likelihoods. In fact, in each run using the emulator, the accumulated execution time of the likelihood codes exceeds the combined run time of gathering training data, training the emulator, and providing the emulated spectra to the sampler. For example, when running an MCMC with the emulator on the $\Lambda$CDM+$\Sigma m_{\nu}$+$N_\mathrm{eff}$ model with the $\mathcal{D}_\mathrm{SPB}$ data set, while requiring a convergence of $R-1<0.01$ with 8 chains using 1 core each, we find that each chain spends about 22 CPUh on the SPT likelihood, 9 CPUh on the Planck likelihood, 1 CPUh on calling \texttt{CLASS} to generate 300-500 spectra, and 1 CPUh on training and evaluating the emulator. Without the emulator, about 90\% of the time is spent on calling \texttt{CLASS}. Thus, the achievable speed-up factor can be roughly approximated by the ratio of the evaluation time of the theory and likelihood code.\footnote{Note that this is only an approximate estimate, not taking into account the fact that the number of calls to the theory code can be lowered by different strategy (like the use of fast-slow parameters), as described in~\cite{Lewis_2013}.} In this work, we find that the exact speed-up factor is model and data-set dependent, and we find speed-up factors of about $10$ to $20$.\footnote{The run times on 1 core are $\sim10$s (\texttt{CLASS}), $\sim0.1$s (SPT, Planck). However, the ratio of nuisance parameters to cosmological parameters is about $3$ to $5$.} In principle, the speed-up factor could be further improved by accelerating the likelihood codes \cite{Prince_2019} or utilizing more efficient sampling algorithms (as done in \cite{bonici2023capsejl}).

For all the models discussed in this work and for the ${\cal D}_\mathrm{SPB}$ data set, we have systematically performed two MCMC runs starting from the same initial conditions (usually, from a guessed covariance matrix providing a poor approximation to the optimal proposal density, especially for the parameter of the extended models). One run called the Boltzmann code systematically for each new model, while the other run used the scheme described in ref.~\cite{günther2023uncertaintyaware} in which calls to the Boltzmann code are gradually replaced by calls to the emulator. For all models, the two runs provided identical contours, means, and confidence limits (after rounding numbers in such a way as to keep two significant digits in the error bars).

\begin{figure}
                \hspace{-2cm} 
                \includegraphics[width=1.2\columnwidth,angle=0]{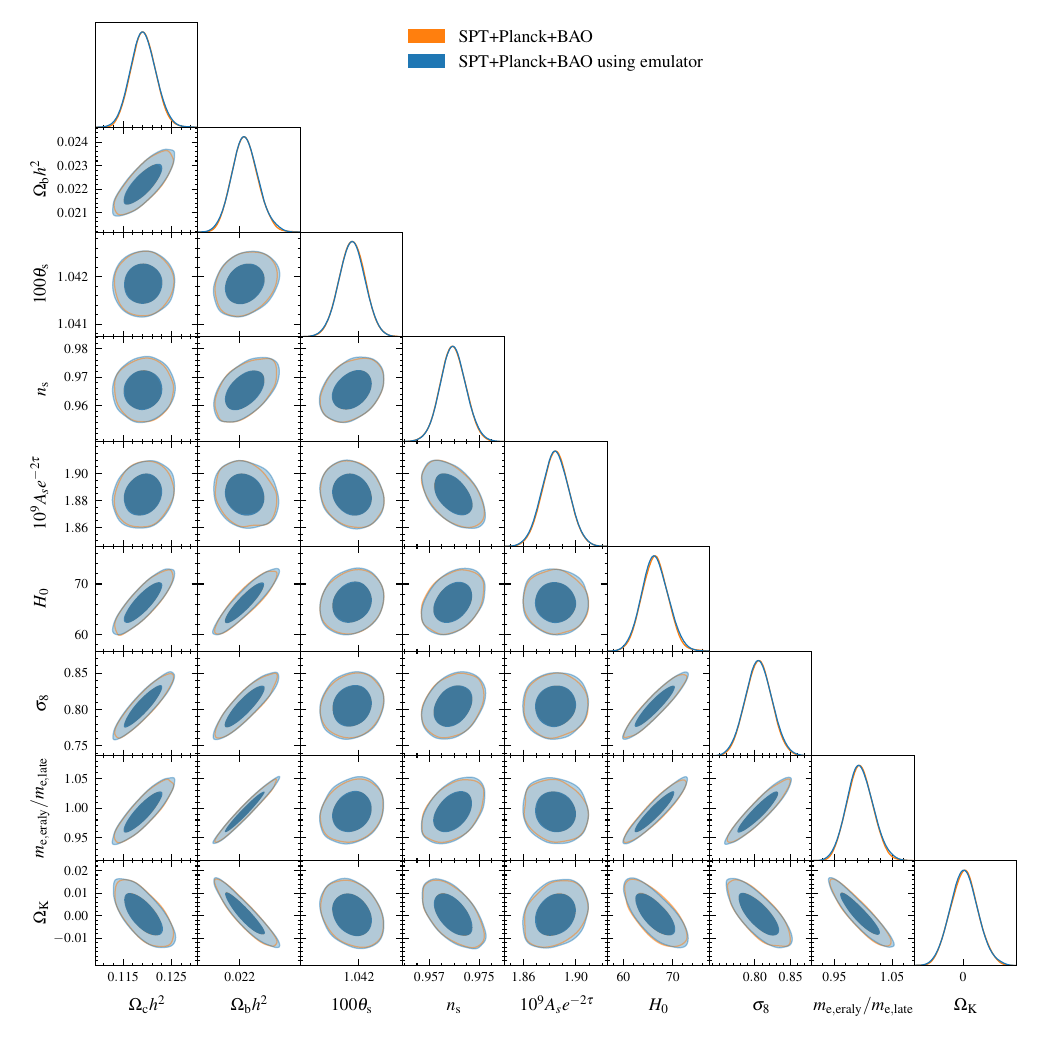}
                \caption{Varying $m_e + \Omega_\mathrm{K}$: 2D contours on relevant cosmological parameters for the default MCMC implementation (orange) and the emulator (blue).}
                \label{fig:me_omk_emulator}
        \end{figure}

In Figure~\ref{fig:me_omk_emulator}, we compare the posteriors for the varying $m_e$+$\Omega_\mathrm{K}$ model using the $\mathcal{D}_{\text{SPBP}}$ data set defined in~\eqref{Eq:Data_sets_name}. The background (orange) lines and contours were obtained with \texttt{Cobaya} running in standard mode, that is, with systematic calls to the Boltzmann code, and the foreground (blue) ones with the emulator mode switched on. We find excellent agreement between the posterior estimates and conclude that the emulator can accurately describe all the relevant effects of this model at the level of CMB power spectra. Note that our method implies that each MCMC builds independently its emulator on the fly. In the present case, for each of our eight chains, the emulator accumulated 400-700 simulations for its own training and was used 60,000-70,000 times to emulate the CMB observables. Thus, by using the emulator, we could avoid $\sim 99\%$ of the simulations (that is, of the calls to \texttt{CLASS}).

\begin{figure}
                \hspace{-2cm} 
                \includegraphics[width=1.2\columnwidth,angle=0]{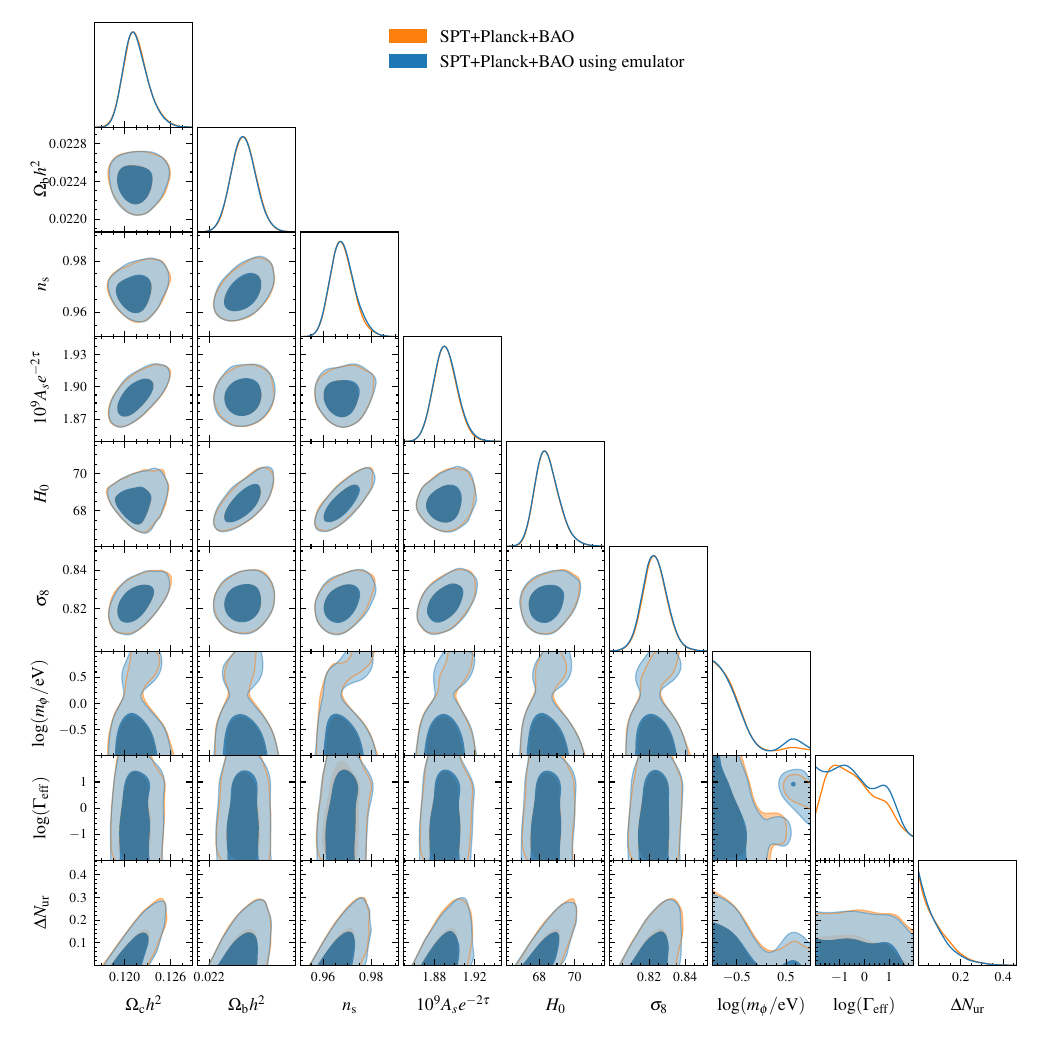}
                \caption{2D contours of the \textit{Majoron} model on relevant cosmological parameters with the default MCMC implementation (orange) and the emulator (blue).}
                \label{fig:maj_emulator}
        \end{figure}

In Figure~\ref{fig:maj_emulator}, we check that there is still an excellent level of agreement between the posterior estimates for the Majoron model using the same data set. This case is particularly challenging because the contours are highly non-Gaussian. Since our emulator assesses its own level of uncertainty, a complicated posterior shape automatically leads to more simulation calls. For the Majoron model, each chain required 1100-1700 \texttt{CLASS} calls and 70,000-90,000 emulator calls.

\subsection{Minimization and best-fit models}
\label{subsec:minimization}

In order to evaluate some of our tension metrics (see section~\ref{sec:metrics}) for a given model and data set, we need to determine the best fit by minimizing the effective $\chi^2=-2 \ln \mathcal{L}$ (where $\mathcal{L}$ is the likelihood). For this purpose, we use the $\texttt{BOBYQA}$~\cite{BOBYQA1,BOBYQA2} algorithm implemented in $\texttt{COBAYA}$. We have set the tolerance parameter (i.e. the parameter that determines the convergence of the minimization, called \texttt{rhoend} in \texttt{BOBYQA}) to be 0.005. Moreover, we perform 104 minimizations for each model to ensure a sufficient exploration of the parameter space to find the global minimum, rather than some random local minimum. We sample the initial points from a Gaussian fit to the posterior distribution for both cosmological and nuisance parameters. 

However, minimization remains a difficult and computationally expensive task. It is further complicated by numerical noise in the Einstein-Boltzmann solver, which originates from ordinary differential equation solvers, from the discrete sampling of integrands, and from the discrete switching of approximation schemes in the code. As a matter of fact, the default precision parameters in Boltzmann codes like {\tt CLASS} or {\tt CAMB} are chosen to have the smallest possible execution time while ensuring unbiased estimates of means and confidence limits in MCMC runs. Since the analysis of MCMCs tends to smooth out random numerical noise, the precision settings are not optimised to remove such noise down to a very small level. To do minimization, it is possible to improve the precision settings of these codes to lower the numerical noise by about one order of magnitude, which is sufficient for obtaining smoother profile likelihoods and removing artificial local maxima created by random noise. However, in this case, the codes become extremely slow, such that performing many minimization is prohibitively expensive. Using an emulator allows us to avoid this expensive solution since, as long as the emulator is trained over a relatively small data set, it smooths over the noise instead of trying to fit it.

By using the emulator previously described~\cite{günther2023uncertaintyaware} to speed up the evaluation of {\tt CLASS} and smooth the numerical noise, we were able to accelerate the minimization task by a factor $\sim \mathcal{O}(10^2)$.\footnote{When doing a single minimization run (rather than Metropolis-Hastings~\cite{Metropilis_Hastings1,Metropolis_Hastings2} MCMCs to find the means and posteriors), we find that the training requires $\mathcal{O}(10^2)-\mathcal{O}(10^3)$ model realizations (e.g. $\sim$100 for $\Lambda \mathrm{CDM}$ and $\sim$1000 for the Majoron model). This is still small compared to the total number of evaluations required by the \texttt{BOBYQA} algorithm, which is of the order of $\mathcal{O}(10^4)-\mathcal{O}(10^5)$.} Note that a similar approach was already used in ref.~\cite{nygaard2023fast}.

Even when using the emulator, the existence of local $\chi^2$ minima can subsist and there is no guarantee that a single minimization converges to the global minimum. Thus, for each model, we still perform $\mathcal{O}(100)$ minimizations starting from random initial conditions (as described above) and we state the minimum after ensuring that a substantial fraction (about 10\%) of found (local) minima accumulate to the same value of $\chi^2$ within a margin of $0.2-0.4$. This level of accuracy is good enough to reach robust conclusions on the DMAP and AIC metrics.

\bibliography{biblio}
\bibliographystyle{utcaps}

\end{document}